\documentclass{article}
\usepackage{amssymb}
\usepackage{amsmath}

\def\nonu{\nonumber}

\def\br{\begin{eqnarray}}
\def\er{\end{eqnarray}}
\def\be{\begin{equation}}
\def\ee{\end{equation}}
\def\0{\nonumber}
\def\lb{\lbrack}
\def\rb{\rbrack}
\def\({\left(}
\def\){\right)}

\relax

\def\a{\alpha}

\def\g{\gamma}

\def\h{ {1\over 2}  }

\def\l{\lambda}

\def\pa{\partial}
\def\pr{\prime}

\def\rlx{\relax\leavevmode}
\def\inbar{\vrule height1.5ex width.4pt depth0pt}
\def\IZ{\rlx\hbox{\sf Z\kern-.4em Z}}
\def\IR{\rlx\hbox{\rm I\kern-.18em R}}
\def\IC{\rlx\hbox{\,$\inbar\kern-.3em{\rm C}$}}
\def\one{\hbox{{1}\kern-.25em\hbox{l}}}

\def\NPB#1#2#3{{\sl Nucl. Phys.} {\bf B#1} (#2) #3}

\def\CMP#1#2#3{{\sl Commun. Math. Phys.} {\bf #1} (#2) #3}

\def\PLA#1#2#3{{\sl Phys. Lett.} {\bf #1A} (#2) #3}
\def\PLB#1#2#3{{\sl Phys. Lett.} {\bf #1B} (#2) #3}

\def\PR#1#2#3{{\sl Phys. Reports} {\bf #1} (#2) #3}

\newtheorem{theorem}{Theorem}

\newtheorem{proposition}[theorem]{Proposition}
\newtheorem{remark}[theorem]{Remark}

\newenvironment{proof}[1][Proof]{\noindent\textbf{#1.} }{\ \rule{0.5em}{0.5em}}
\input{tcilatex}
\begin{document}

\title{\textbf{ Super WZNW with Reductions to Supersymmetric and Fermionic Integrable Models}}
\author{\textit{\thanks{%
jfg@ift.unesp.br}J. F. Gomes}, \textit{\thanks{%
david@ift.unesp.br}D. M. Schmidtt} and \thanks{%
zimerman@ift.unesp.br}\textit{A. H. Zimerman} \\
\\
\textit{Instituto de F\'{\i}sica Te\'{o}rica-IFT/UNESP}\\
\textit{Rua Pamplona 145, CEP 01405-000, S\~{a}oPaulo-SP, Brasil.}}
\maketitle

\begin{abstract}
A systematic construction for an  action describing  a class of supersymmetric integrable models 
 as well as for pure fermionic theories
is discussed in terms of the gauged WZNW model associated to twisted affine Kac-Moody algebras.  
Explicit examples  of the $N=1,2$ super sinh(sine)-Gordon models are discussed in detail. 
Pure fermionic theories arises for cosets $sl(p,1)/sl(p)\otimes u(1)$ when  a  maximal kernel condition is fulfilled.
The integrability condition for such models is discussed and it is shown that 
 the simplest example 
when $p=2$ leads to the constrained Bukhvostov-Lipatov, Thirring, scalar massive and pseudo-scalar massless Gross-Neveu models.

\end{abstract}

\section{Introduction}

The construction and classification  of  integrable models is 
known to  be underlined by an affine  Lie  algebraic structure.  
Moreover, a systematic construction of soliton solutions may be 
understood in terms of representation theory of  affine Lie algebras.
The action of several relativistic integrable models  can be derived  from  
reductions of the Wess-Zumino-Novikov-Witten (WZNW) model representing 
a 2D field theory in a group manifold.  For the finite dimensional Lie algebra, 
the formulation of the Hamiltonian reduction involving Kac-Moody currents (one-loop) 
was proposed in \cite{ora}.
In ref. \cite{twoloop}  
a generalization of WZNW model with currents satisfying  a two-loop 
Kac-Moody  algebra was proposed and    its reduction lead to the construction 
of    conformal affine Toda models.  A gauged two-loop 
WZNW version was considered in \cite{dyonic} in order to derive a 
class of dyonic ($U(1)$) integrable models and was further generalized 
for non abelian internal structure in \cite{isospin}.

The construction of supersymmetric  integrable models of the sinh-Gordon (mKdV)-type was 
proposed in \cite{half-integer gradations} based upon the Riemann-Hilbert (RH) factorization problem. 
The object of this paper is to  propose an alternative and complementary  construction 
based upon the  gauged two-loop WZNW action for   affine super Lie algebras.  
The key ingredient is the decomposition  of a twisted affine structure into integer and 
semi-integer subspaces followed by 
a Gauss-type  parametrization which, in turn,  defines  the bosons and the fermions of the theory.
Our construction uncovers  the  group theoretic origin of the physical fields within which the 
fermions parametrize  grade $\pm 1/2$ subspace , $W_{\pm 1/2}\in {\mathfrak{g}} _{\pm 1/2}$ whilst  
the bosons parametrize the zero grade subspace ${\mathfrak{g}}_0$.  A local supersymmetry condition 
found in ref. \cite{half-integer gradations} naturaly arises within the formalism and   
 is responsible for the truncation of the potential.
  Explicit examples 
for the twisted affine super algebras $\hat {sl}(2,1)$ and $\hat {sl}(2,2)$  yields the 
$N=1$ and $N=2$ super sinh-Gordon models.  
The implementation of the supersymmetry condition  naturaly leads to constructing   
the model associated to grade two of  the higher grading generalization of the Toda 
 systems coupled to matter   proposed in  \cite{gervais-saveliev} , \cite{ Ferreira} 
 in terms  of zero curvature representation. 
 
 By a suitable   decomposition of the $\hat{sl}(p,1)$ affine twisted super algebra, 
 a structure with maximal kernel within the bosonic zero grade sector can be constructed. 
  The whole bosonic subalgebra, namely $sl(p)\otimes U(1)$ may be  gauged away  by 
  constructing  a subsequent gauged WZNW action  similarly to  one of ref. 
  \cite{isospin} resulting in a pure fermionic  model. We  develop explicitly 
  the $sl(2,1)$ case and discuss its integrability.  Fermionic integrable 
  models like constrained Bukhvostov-Lipatov, Thirring, scalar massive and pseudo-scalar massless Gross-Neveu models  
   may be generated from this formalism.

This paper is organized as follows. Section 2 is devoted to 
review  the construction of the Lax pair of the Leznov-Saveliev (LS) equations by solving
the  RH factorization problem by means of the dressing technique. 
In  Section 3  we discuss the  Hamiltonian reduction of a two-loop WZNW. The main
ingredients is the introduction of auxiliary gauge fields and a Gauss-type decomposition 
of the super group element. 
Invariance of the gauged WZNW action allows the reduction of the degrees 
of freedom leading to effective  coset elements within the $0, \pm 1/2 $ graded subspaces ,
${\mathfrak{g}}_0$ and  ${\mathfrak{g}}_{\pm 1/2}$, parametrized by  the bosons and fermions of the theory respectively. 
The effective action simplifyies considerably  after  imposing a subsequent 
constraint $Q_{0 \pm}^{(2)}=0$ already considered in ref. \cite{half-integer gradations} within 
the locality of supersymmetry transformation.


Section 4 is devoted to the explicit construction of examples involving the supersymmetric $N=1$ and $N=2$ sinh (sine)-Gordon  models. 
In Section 5 we discuss the extreme cases where the bosonic Kernel is maximal.  
The basic prototype  is given by the coset  $sl(p,1)/sl(p) \otimes U(1)$ which 
generates pure fermionic theories.    
The integrability of these theories is discussed and the simplest example 
when $p=2$ is considerd in detail.
Models like the constrained Bukhvostov-Lipatov, Thirring, scalar massive and pseudo-scalar massless Gross-Neveu models  are shown to belong 
to such class  related to $sl(2,1)/sl(2) \otimes U(1)$.

\section{The extended Riemann-Hilbert factorization problem and the Dressing
formalism.}

Here we briefly review the algebraic approach given in \cite{Symmetry
flows...} where the dressing formalism was used to unify symmetry
flows (Isospectral and Non-abelian) of Integrable Hierarchies related to $%
\mathbb{Z}
$-graded Affine Lie algebras. These ideas were applied to Twisted Affine Lie
superalgebras in \cite{half-integer gradations} with a $%
\mathbb{Z}
/2$ gradation were the Lax operators for several supersymmetric integrable
Hierarchies were derived by solving an extended Riemann-Hilbert
factorization problem.

\subsection{Symmetry flows and Isospectral times.}

The Riemann-Hilbert factorization problem allows us to define
an Integrable structure and a related hierarchy of non-linear partial
differential equations. This also provides the explicit form of the Lax 
 operators from which the equations of motion can be written as a zero
curvature representation.

 Consider a Twisted Loop superalgebra decomposed  by a grading
operator $Q$ into $%
\mathbb{Z}
/2$-graded spaces satisfying $\left[ Q,\hat {\mathfrak{g}}_{i}\right] =i%
\hat {\mathfrak{g}}_{i}$, $i\in 
\mathbb{Z}
/2$ such that $\hat{\mathfrak{g}}=\underset{i\in 
\mathbb{Z}
/2=-\infty }{\overset{+\infty }{\dbigoplus }}\hat {\mathfrak{g}}_{i}$.  Let 
 $E^{(n)}$ be  a constant Bosonic semisimple element  of degree $n$
which induces the following algebra decomposition $\hat{\mathfrak{g}}=%
\mathcal{K}\tbigoplus \mathcal{M}$, where the spaces $\mathcal{K}$ and $\mathcal{M}
$ are the kernel and the image of the adjoint operator $adE^{(n)}\left( \ast
\right) \equiv \left[ E^{(n)},\ast \right] $ respectively.

The integrable structure treated here is derived from an extended
Riemann-Hilbert factorization problem involving positive and negative time
flows%
\begin{equation}
\exp \left[ -\sum_{n=1}^{\infty }E^{(n)}t_{n}\right] g_{0}\exp \left[
\sum_{n=1}^{\infty }E^{(-n)}t_{-n}\right] =\Theta ^{-1}(t)\Pi (t)\text{ ,}
\label{R-H problem}
\end{equation}%
where $g_{0}$ is a constant element in $\widetilde{G}$ and the dressing
matrices are given by the following exponentials
\begin{eqnarray*}
\Theta (t) &=&\exp \left( \sum_{i\in 
\mathbb{Z}
/2=1/2}^{+\infty }W_{-i}(t)\right) \text{ ,\ }\Pi (t)=B(t)M(t) \\
\text{ }M(t) &=&\exp \left( \sum_{i\in 
\mathbb{Z}
/2=1/2}^{+\infty }W_{+i}(t)\right) \text{\ , }B(t)=\exp \left( \hat {%
\mathfrak{g}}_{0}\right) .
\end{eqnarray*}

From (\ref{R-H problem}) we get the flow equations%
\begin{eqnarray}
\frac{\partial }{\partial t_{n}}\Theta (t) &=&\left( \Theta E^{(n)}\Theta
^{-1}\right) _{-}\Theta (t)\text{ , }\frac{\partial }{\partial t_{n}}\Pi
(t)=-\left( \Theta E^{(n)}\Theta ^{-1}\right) _{+}\Pi (t)\text{ \ }
\label{1 relation R-H} \\
\text{\ }\frac{\partial }{\partial t_{-n}}\Theta (t) &=&-\left( \Pi
E^{(-n)}\Pi ^{-1}\right) _{-}\Theta (t)\text{\ ,\ }\frac{\partial }{\partial
t_{-n}}\Pi (t)=\left( \Pi E^{(-n)}\Pi ^{-1}\right) _{+}\Pi (t),\nonu \\
\label{2 relation R-H}
\end{eqnarray}%
where $\left( \ast \right) _{\pm }$ denotes projection on the $\geq 0$ and $<0$
grades respectively. Taking $n=1$ for the
first equation in (\ref{1 relation R-H}) we get $(E^{(+1)}\equiv E_{+})$%
\begin{equation*}
\frac{\partial }{\partial t_{1}}\Theta (t)=\text{ }\Theta (t)E_+ -\left(
E_+ +A_{0}+A_{1/2}+Q_{0+}^{(2)}\right) \Theta (t)\text{\ ,}
\end{equation*}%
where $A_{0}\equiv \left[ W_{-1},E_{+}\right] \in \mathcal{M}$ , $%
A_{1/2}\equiv \left[ W_{-1/2},E_{+}\right] \in $ $\mathcal{M}$ and $%
Q_{0+}^{(2)}\equiv \frac{1}{2}\left[ W_{-1/2},\left[ W_{-1/2},E_{+}\right] %
\right] \in \mathcal{K}\tbigoplus \mathcal{M}$.\ From this equation we find
the dressing relation%
\br
L_{1} &=&\Theta L_{1}^{V}\Theta ^{-1} \nonu \\
L_{1} &=&\frac{\partial }{\partial t_{1}}+E_{+}^{\text{ }%
}+A_{0}+A_{1/2}+Q_{0+}^{(2)} \label{a} \\
L_{1}^{V} &=&\frac{\partial }{\partial t_{1}}+E_{+}^{\text{ }},\nonu
\er
which relates the Lax operator $L_{1}$ and the vacuum Lax operator $%
L_{1}^{V} $ by means of a Dressing gauge transformation. Similarly, by
considering the negative flow $n=1$ for the first equation in (\ref{2
relation R-H}) we get $(E^{(-1)}\equiv E_{-})$ 
\br
L_{-1} &=&\Theta L_{-1}^{V}\Theta ^{-1} \nonu \\
L_{-1} &=&\frac{\partial }{\partial t_{-1}}+B\left( E_{-}+J_{-1/2}\right)
B^{-1} \label{b} \\
L_{-1}^{V} &=&\frac{\partial }{\partial t_{-1}}+E_{-}^{\text{ }}\nonu 
\er
where $J_{-1/2}\equiv -\left[ W_{+1/2},E_{-}\right] $ $\in \mathcal{M}$.

The additional term $Q_{0+}^{(2)}$ appearing in the Lax pair comes entirely
from the $-1/2$ grade component of the dressing matrix $\Theta $ and when different
from zero gives rise to Integrable models with non-local supersymmetry c.f 
\cite{half-integer gradations}. Local supersymmetry transformations are obtained 
when\textit{\ } 
\begin{equation}
Q_{0 + }^{(2)}\equiv \frac{1}{2}ad^{2}W_{- 1/2}\left( E_{+ }\right) = \frac{1}{2}\left[ W_{-1/2},\left[ W_{-1/2},E_{+}\right] %
\right] = 0%
\text{.}  \label{Locality in R-H}
\end{equation}
The algebra of symmetries of the model is identified with a centralizer of
the generator of isospectral deformations, namely $E_{+}$. In this setting,
we will associate to all positive grade elements $K_{i}^{+}\subset $ $%
\mathcal{K}$ , the following transformation equations%
\begin{eqnarray*}
\delta _{K_{i}}\Theta (t) &=&\left( \Theta K_{i}\Theta ^{-1}\right)
_{-}\Theta (t)\text{ } \\
\text{\ }\delta _{K_{i}}\Pi (t) &=&-\left( \Theta K_{i}\Theta ^{-1}\right)
_{+}\Pi (t)\text{.}
\end{eqnarray*}
The map defined by $K_{i}\rightarrow \delta _{K_{i}}$ defines the
Homomorphism $\left[ \delta _{K_{i}},\delta _{K_{i}}\right] \Theta =\delta _{%
\left[ K_{i},K_{j}\right] }\Theta $ and due to the fact that $K_{i}^{+}$
belongs to the kernel $\mathcal{K}$ of the operator $adE_{\pm }\left( \ast
\right) \equiv \left[ E_{\pm },\ast \right] $ they commute with the
isospectral deformations and are symmetries of the Integrable model. The
existence of a non-trivial Fermionic kernel for the operators $adE_{\pm
}(\ast )$ will lead to the existence of  half integer  flows and therefore 
to supersymmetry transformations. 
Given a  grade one-half element,  $D^{1/2}\in $ $\mathcal{K}$, define
 a Fermionic flow%
\begin{equation}
\frac{\partial }{\partial t_{1/2}}\Theta (t)\equiv \delta _{D^{1/2}}\Theta
(t)=\left( \Theta D^{1/2}\Theta ^{-1}\right) _{-}\Theta (t)\text{ }
\label{fermionic flow}
\end{equation}%
from which we get  the  dressing expression%
\begin{eqnarray*}
L_{+1/2} &=&\Theta L_{+1/2}^{V}\text{ }\Theta ^{-1} \\
L_{+1/2} &=&\left( \frac{\partial }{\partial t_{1/2}}+D^{1/2}+D^{0}\right) \\
L_{+1/2}^{V} &=&\left( \frac{\partial }{\partial t_{1/2}}+D^{1/2}\right)
\end{eqnarray*}%
where $D^{0}\equiv \left[ W_{-1/2},D^{1/2}\right] .$ This flow commutes with
the ones generated by $\frac{\partial }{\partial t_{1}}$ and $\frac{\partial 
}{\partial t_{-1}}$ and the relations 
\begin{eqnarray}
\left[ L_{+1/2},L_{-1}\right] &=&0  \label{L1/2,L-1} \\
\left[ L_{+1/2},L_{+1}\right] &=&0  \notag \\
\left[ L_{+1},L_{-1}\right] &=&0  \notag
\end{eqnarray}%
follows as   compatibility equations which guarantees the invariance under
supersymmetry transformations of the equations of motion written as  zero curvature representation , i.e.,
$\left[ L_{+1},L_{-1}\right] =0.$

The grade -1 components of the first and third equations  (\ref{L1/2,L-1})
allows to write%
\begin{equation*}
A_{0}=-\frac{\partial }{\partial t_{1}}BB^{-1},\text{ \ }D^{0}=-\frac{%
\partial }{\partial t_{1/2}}BB^{-1}\text{ ,}
\end{equation*}%
from which we extract the supersymmetry transformations among the matrix
fields%
\begin{eqnarray}
\partial _{1/2}J_{-1/2} &=&\left[ E_{-},B^{-1}D^{1/2}B\right]
\label{SUSY transformations} \\
\partial _{-1}\left( \partial _{1/2}BB^{-1}\right) &=&\left[
BJ_{-1/2}B^{-1},D^{1/2}\right]  \notag \\
\left[ E_{+},\partial _{1/2}BB^{-1}\right] &=&\left[ A_{1/2},D^{1/2}\right] 
\notag \\
\partial _{1/2}A_{1/2} &=&\left[ \partial _{1/2}BB^{-1},A_{1/2}\right] -%
\left[ \partial _{+1}BB^{-1},D^{1/2}\right] .  \notag
\end{eqnarray}%
From this analysis we obtain the Super-Integrable Leznov-Saveliev equations (see \cite{half-integer gradations} )
\begin{eqnarray}
\partial _{-1}A_{1/2} &=&\left[ E_{+},BJ_{-1/2}B^{-1}\right]
\label{SUSY equations} \\
\partial _{-1}\left( \partial _{+1}BB^{-1}\right) &=&\left[
BE_{-}B^{-1},E_{+}\right] +\left[ BJ_{-1/2}B^{-1},A_{1/2}\right]  \notag \\
\partial _{+1}J_{-1/2} &=&\left[ E_{-},B^{-1}A_{1/2}B\right] \text{.}  \notag
\end{eqnarray}%
together its Lax pair%
\begin{eqnarray}
L_{+1} &=&\partial _{+1}-\partial _{+1}BB^{-1}+E_{+}^{\text{ }}+A_{1/2}
\label{Lax pair} \\
L_{-1} &=&\partial _{-1}+B\left( E_{-}+J_{-1/2}\right) B^{-1}.  \label{23}
\end{eqnarray}

The  relativistic hierarchy ($t_{-1}$-flow, $t_{-1} = z, t_1 = \bar z$) given by equations (\ref{SUSY equations}) can
also be formulated as  Hamiltonian reduction procedure of a 2-Loop WZNW
model by imposing an infinite number of first class constraints, as we will
now show.

\section{Supersymmetric Affine Toda Field Theories $s$-$ATFT$.}

We now show how the action governing the equations of motion of the
Supersymmetric Affine Toda models can be derived from a gauged 2- loop WZNW
action where the fundamental matrix fields takes values on a Twisted Affine
Lie superalgebra.

\subsection{Hamiltonian reduction of the 2- loop WZNW model.}

We assume the existence of a \textit{Twisted Affine Lie Superalgebra} (see for instance  \cite%
{sorba}) $\mathfrak{g}$ endowed with a Half-integer gradation according
to a grading operator $Q$ . We also identify the $%
\mathbb{Z}
_{2}$ statistics of the fields with the $%
\mathbb{Z}
_{2}$ structure of the classical superalgebra $\widehat{\mathfrak{g}}$. This means
that Bosonic-Fermionic fields will parametrize respectively the even-odd
part of the superalgebra with $\Phi _{B}\in \digamma \otimes \mathfrak{g}%
_{B} $ and $\Psi _{F}\in \digamma \otimes \mathfrak{g}_{F}$ given by 
\begin{equation}
\Phi _{B}\equiv \phi _{i}\otimes \mathfrak{g}_{B}^{i}\text{ , }\Psi
_{F}\equiv \psi _{k}\otimes \mathfrak{g}_{F}^{k}\text{ ,}  \label{sections}
\end{equation}%
where $\digamma $ is the field space depending on $t_{\pm 1}.$

The structure of the twisted affine  superalgebra  $\widehat{\mathfrak{g}}$ is then  given by
\br \widehat{\mathfrak{g}}=\underset{i\in 
\mathbb{Z}
/2=-\infty }{\overset{+\infty }{\dbigoplus }}\widehat{\mathfrak{g}}_{i}, \qquad \left[ Q,%
\widehat{\mathfrak{g}}_{i}\right] =i\widehat{\mathfrak{g}}_{i}
\er
Let us consider a group element  expressed in a generalized Gauss-type
 form  as follows%
\begin{eqnarray}
g =K_{<}\;\; \Gamma \;\; K_{>}  \qquad 
\Gamma =\Phi B\Psi ,  \label{parametrization}
\end{eqnarray}%
where 
\br 
K_{<}\in \exp \( \widehat{\mathfrak{g}}_{\leq -1}\) , \qquad  K_{>}\in \exp \(\widehat{\mathfrak{g}}_{\geq 1}\),
\er
 \br
 \Phi =\exp
(W_{-1/2}),\qquad  B\in \exp \(\widehat{\mathfrak{g}}_{0}\), \qquad \Psi =\exp (W_{+1/2})
\label{field}
\er 
and $W_{\pm 1/2} \in \widehat{\mathfrak{g}}_{\pm 1/2}$.
Propose  the gauged WZNW action 
\br
S[g,A_{\pm }] &=&S_{WZNW}[g]- \nonu \\
&-&\frac{k}{2\pi }\int \left\langle A_{-}\left( \partial
_{+}gg^{-1}-E_{+}\right) +A_{+}\left( g^{-1}\partial _{-}g-E_{-}\right)
+A_{-}gA_{+}g^{-1}\right\rangle ,\nonu \\
\label{action1}
\er
which is invariant under the local gauge transformation 
\br
g &\rightarrow &g^{\prime }=\alpha g\beta \text{ },\text{ }\alpha \in 
\exp \( \widehat{\mathfrak{g}}_{\leq -1}\) \text{ and }\beta \in \exp \(\widehat{\mathfrak{g}}_{\geq 1}\) \nonu \\
A_{+}^{\prime } &=&\beta ^{-1}A_{+}\beta +\partial _{+}\beta ^{-1}\beta \nonu \\
A_{-}^{\prime } &=&\alpha A_{-}\alpha ^{-1}+\alpha \partial _{-}\alpha ^{-1},
\label{transf1}
\er
where $A_{+}\in \widehat{\mathfrak{g}}_{\geq +1}$ , $A_{-}\in \widehat{%
\mathfrak{g}}_{\leq -1}$ and $\left\langle \ast \right\rangle $ is the
generalized supertrace, i.e take the supertrace followed by projection using
the orthogonality condition $\left\langle \widehat{\mathfrak{g}}_{i}\widehat{%
\mathfrak{g}}_{j}\right\rangle =a$ $\delta _{i+j,0}.$ 
Since the action (\ref{action1}) is invariant under (\ref{transf1}) we may choose 
$\alpha
=K_{<}^{-1}$ and $\beta =K_{>}^{-1}$ which in practice amounts to the elimination of an
infinite number of fields (first class constraints)  ending up with
\begin{eqnarray*}
S[\Gamma ,A_{\pm }^{\prime }] &=&S_{WZNW}[\Gamma ]- \\
&&-\frac{k}{2\pi }\int \left\langle A_{-}^{\prime }\left( \partial
_{+}\Gamma \Gamma ^{-1}-E_{+}\right) +A_{+}^{\prime }\left( \Gamma
^{-1}\partial _{-}\Gamma -E_{-}\right) +A_{-}^{\prime }\Gamma A_{+}^{\prime
}\Gamma ^{-1}\right\rangle .
\end{eqnarray*}
For the action $S_{WZNW}[\Gamma ]$ above, we use the Polyakov-Wiegmann
identity\footnote{%
This is given by:
\br
S_{WZNW}[ABC] &=&S_{WZNW}[A]+S_{WZNW}[B]+S_{WZNW}[C] \nonu \\
&-&\frac{k}{2\pi }\int \left\langle 
\left( A^{-1}\partial _{-}A\right) \left( \partial _{+}BB^{-1}\right)
+\left( B^{-1}\partial _{-}B\right) \left( \partial _{+}CC^{-1}\right) \right\rangle \nonu \\
&-& \frac{k}{2\pi }\int \left\langle  \left( A^{-1}\partial _{-}A\right) B\left( \partial _{+}CC^{-1}\right)
B^{-1}%
\right\rangle ,
\er
where
\begin{equation*}
S_{WZNW}[g]=-\frac{k}{4\pi }\int \left\langle \left( g^{-1}\partial
_{+}g\right) \left( g^{-1}\partial _{-}g\right) \right\rangle +\frac{k}{%
12\pi }\int \epsilon ^{IJK}\left\langle \left( g^{-1}\partial _{I}g\right)
\left( g^{-1}\partial _{J}g\right) \left( g^{-1}\partial _{K}g\right)
\right\rangle
\end{equation*}}
to decompose it as%
\begin{equation}
S_{WZNW}[\Gamma ]=S_{WZNW}[B]-\frac{k}{2\pi }\int \left\langle \left( \Phi
^{-1}\partial _{-}\Phi \right) B\left( \partial _{+}\Psi \Psi ^{-1}\right)
B^{-1}\right\rangle \text{ }  \label{PW}
\end{equation}%
and for the second term we seek  for the non-zero contributions of the inner
product%
\begin{equation*}
I=\left\langle A_{-}\left( \partial _{+}\Gamma \Gamma ^{-1}-E_{+}\right)
+A_{+}\left( \Gamma ^{-1}\partial _{-}\Gamma -E_{-}\right) +A_{-}\Gamma
A_{+}\Gamma ^{-1}\right\rangle 
\label{gaugeterms}.
\end{equation*}
Analyzing term by term we have
\begin{eqnarray*}
\left\langle A_{-}\left( \left( \partial _{+}\Gamma \Gamma ^{-1}\right)
|_{\geq +1}-E_{+}\right) \right\rangle &=&\left\langle A_{-}\left( \Phi
B\partial _{+}\Psi \Psi ^{-1}B^{-1}\Phi ^{-1}-E_{+}\right) \right\rangle \\
\left\langle A_{+}\left( \left( \Gamma ^{-1}\partial _{-}\Gamma \right)
|_{\leq -1}-E_{-}\right) \right\rangle &=&\left\langle A_{+}\left( \Psi
^{-1}B^{-1}\Phi ^{-1}\partial _{-}\Phi B\Psi -E_{-}\right) \right\rangle .
\end{eqnarray*}%
Solving now  the  equations of motion for the  gauge fields,
\begin{eqnarray*}
A_{+} &=&\Gamma ^{-1}E_{+}\Gamma -\Psi ^{-1}\partial _{+}\Psi \\
A_{-} &=&\Gamma E_{-}\Gamma ^{-1}-\partial _{-}\Phi \Phi ^{-1}
\end{eqnarray*}%
and  plugging them back in (\ref{gaugeterms}) yields the effective action,
\begin{eqnarray}
I_{eff}&=&-\left\langle \left( \Phi ^{-1}E_{+}\Phi \right) B\left( \Psi E_{-}\Psi
^{-1}\right) B^{-1}\right\rangle -\left\langle \left( \Phi ^{-1}\partial
_{-}\Phi \right) B\left( \partial _{+}\Psi \Psi ^{-1}\right)
B^{-1}\right\rangle   \notag \\
&&+\left\langle \left( \Phi ^{-1}E_{+}\Phi \right) \left( \Phi ^{-1}\partial
_{-}\Phi \right) +\left( \Psi E_{-}\Psi ^{-1}\right) \left( \partial
_{+}\Psi \Psi ^{-1}\right) \right\rangle .
\label{gauge terms}
\end{eqnarray}

The term involving the two derivatives in (\ref{PW}) and (\ref%
{gauge terms}) cancel each other leaving only linear terms in the
derivatives which are the correct kinetic terms expected for Fermionic
fields. Finally, we have%
\begin{eqnarray}
S_{eff}[\Phi ,B,\Psi ] &=&S_{WZNW}[B]+\frac{k}{2\pi }\int \left\langle \left( \Phi
^{-1}E_{+}\Phi \right) B\left( \Psi E_{-}\Psi ^{-1}\right)
B^{-1}\right\rangle  \nonu \\
&&-\frac{k}{2\pi }\int \left\langle \left( \Phi ^{-1}E_{+}\Phi \right)
\left( \Phi ^{-1}\partial _{-}\Phi \right) +\left( \Psi E_{-}\Psi
^{-1}\right) \left( \partial _{+}\Psi \Psi ^{-1}\right) \right\rangle \text{.%
}\nonu \\
\textbf{} \label{full action} 
\end{eqnarray}
The effective action (\ref{full action})  possess, in principle, 
an infinite number of terms due to the Baker-Haussdorff expansion in the potential.  
Although these terms  may eventually truncate  due to either, the Grassmaniann 
 character of the fields or to the Nilpotency of the  subspaces involved, 
 other cases of  interest may be obtained by reduction as we shall now discuss.

\subsection{The case  where $ Q_{0\pm } = 0$ and the Supersymmetric Affine Toda Field Theories }

Now, guided by (\ref{Locality in R-H}) and the symmetric structure of (\ref{full action}), 
 we propose  the following  subsidiary conditions,
 \br
Q_{0+ }^{(2)}&=& \left( \Phi ^{-1}E_{+}\Phi \right) |_{0} = \frac{1}{2}\left[ W_{-1/2},\left[ W_{-1/2},E_{+}\right] %
\right] = 0, \nonu \\
Q_{0- }^{(2)}&=& \left( \Psi E_{-}\Psi^{-1} \right) |_{0} = \frac{1}{2}\left[ W_{1/2},\left[ W_{1/2},E_{-}\right] %
\right] = 0, 
\label{SUSY-local conditions}
\er
which  ensures  finite number of terms 
coming from the Baker-Haussdorf expansions of  $\left( \Phi
^{-1}E_{+}\Phi \right) $ and $\left( \Psi E_{-}\Psi ^{-1}\right) $ appearing
in the effective   action (\ref{full action}).  
Assuming (\ref{SUSY-local conditions})   
we write the action (\ref{full action}) in terms of fields  given in (\ref{field}) as,
\begin{eqnarray}
S_{Toda}[B,W_{\pm 1/2}] &=&S_{WZNW}[B]-\frac{k}{2\pi }\int \left\langle \left( %
\left[ W_{-1/2},E_{+}\right] B\left[ W_{+1/2},E_{-}\right] B^{-1}\right)
\right\rangle  \nonu \\ 
&&-\frac{k}{4\pi }\int \left\langle
\partial _{-}W_{-1/2}\left[ W_{-1/2},E_{+}\right] -\partial _{+}W_{+1/2}%
\left[ W_{+1/2},E_{-}\right] \right\rangle  \nonu \\
&&+\frac{k}{2\pi }\int \left\langle \left( E_{+}BE_{-}B^{-1}\right)\right\rangle 
\label{SUSY-Toda action}
\end{eqnarray}

An arbitrary variation of this action, i.e.,  
\begin{eqnarray*}
\frac{2\pi }{k}\delta S_{Toda} &=&\dint \left\langle \left( \delta
BB^{-1}\right) \left\{ \partial _{-}\left( \partial _{+}BB^{-1}\right) -%
\left[ E_{+},BE_{-}B^{-1}\right] -\left[ A_{1/2},BJ_{-1/2}B^{-1}\right]
\right\} \right\rangle  \\
&+&\int \left\langle \delta W_{+1/2}\left\{ \partial _{+}J_{-1/2}-\left[
E_{-},B^{-1}A_{1/2}B\right] \right\} \right\rangle \nonu \\
&+&\int \left\langle \delta
W_{-1/2}\left\{ \partial _{-}A_{1/2}+\left[ E_{+},BJ_{-1/2}B^{-1}\right]
\right\} \right\rangle ,
\end{eqnarray*}%
yields the following equations of motion 
\begin{eqnarray}
\partial _{-}A_{1/2} &=&\left[ BJ_{-1/2}B^{-1},E_{+}\right]  \notag \\
\partial _{-}\left( \partial _{+}BB^{-1}\right) &=&\left[ E_{+},BE_{-}B^{-1}%
\right] +\left[ A_{1/2},BJ_{-1/2}B^{-1}\right]  \notag \\
\partial _{+}J_{-1/2} &=&\left[ E_{-},B^{-1}A_{1/2}B\right] \text{ .}
\label{SUSY-Toda equations}
\end{eqnarray}
where 
\begin{eqnarray}
A_{1/2} \equiv \left[ W_{-1/2},E_{+}\right], \qquad   
J_{-1/2} \equiv -\left[ W_{+1/2},E_{-}\right] . \label{fields conection} 
\end{eqnarray}
Equations  (\ref{SUSY-Toda equations}) coincide  precisely with  the $LS$ found before  
(see (\ref{SUSY equations})) if we identify the flows $\frac{\partial }{\partial t_{\pm 1}}=\partial _{\pm
1}\rightarrow \pm \partial _{\pm }$ .

The integrability of the model can be shown from  the Lax operators 
\br
L_{+1} = \pa_+ - \pa_+BB^{-1} +E_+ + A_{1/2},\qquad 
L_{-1} = \pa_- - B(E_- + J_{-1/2})B^{-1} 
\label{lax}
\er 
where the  equations of motion (\ref{SUSY-Toda equations}) are written in the zero curvature representation 
\br
\left[ L_{+1}, L_{-1}\right] = 0
\label{zcc}
\er
  A natural question that arises  at this point is whether 
Supersymmetry transformations (\ref{SUSY transformations}) keeps the action (\ref{SUSY-Toda action})
unchanged.   Such proof   is given in appendix A where it is shown that (\ref{SUSY-Toda action}) is in fact invariant under the supersymmetry transformation
\begin{eqnarray}
\partial _{1/2}J_{-1/2} &=&\left[ E_{-},B^{-1}D^{1/2}B\right]
\label{susy transformations} \\
\partial _{1/2}BB^{-1} &=&\left[ D^{1/2},W_{-1/2}\right]  \notag \\
\partial _{1/2}A_{1/2} &=&\left[ \left[ D^{1/2},W_{-1/2}\right] ,A_{1/2}%
\right] -\left[ \partial _{+}BB^{-1},D^{1/2}\right].
\label{supp}
\end{eqnarray}
Notice that equations (\ref{SUSY-Toda equations}) are of the same form as those 
 introduced in \cite{gervais-saveliev} (eqns. (2.12) - (2.14)) within the construction of $M=2$ higher grading Toda field theories 
 coupled to  matter, namely,
 \br
 \partial _{-}(g_0 \pa_+W_1^+ g_0^{-1}) &=&\left[ E_{+}, \pa_- W_1^- \right], \nonu \\
 \pa_+ (g_0^{-1} \pa_- g_0) &=& \lb E_-, g_0^{-1} E_+ g_0\rb + \lb g_0^{-1} \pa_- W_1^- g_0, \pa_+ W_1^+\rb, \nonu \\
 \partial _{+}g_0^{-1} \pa_- W_1^- g_0  &=&\left[ E_{-},\pa_+ W_1^-\right]
 \label{ger}
 \er
 After the change of variables  ( and $\pa_- \rightarrow -\pa_-$) 
 \br
 A_{1/2} = \lb W_{-1/2}, E_+ \rb \rightarrow g_0 \pa_+ W_1^+ g_0^{-1}, \nonu \\
 J_{-1/2} = -\lb W_{1/2}, E_- \rb \rightarrow g_0^{-1} \pa_- W_1^- g_0
 \label{variables}
 \er
 we find that eqns. (\ref{ger}) 
 coincide precisely with our eqns. (\ref{SUSY-Toda equations}) if we identify $B\rightarrow g_0$.
 Although their fields $W_1^{\pm}$   were Bosonic and $ \mathbb{Z} $-gradations 
 were considered, their action coincide precisely with (\ref{SUSY-Toda action})
 with the difference that no analogue for the
subsidiady conditions  (\ref{SUSY-local conditions}) was considered. 
Here the matter fields are
Fermionic and are introduced within  a group theoretic background
 by $W_{\pm 1/2}.$ We also emphasize on the locality of
the action (\ref{SUSY-Toda action}) in constrast to what has been argued in
the literature \cite{gervais-saveliev}, \cite{Ferreira}, in the sense that
the connection between the action and the equations of motion involves a
non-local field transformation as in (\ref{variables}).

 \section{Examples  Super sinh-Gordon Type Models}
 
 \subsection{The $N=1$ $\hat {sl}(2,1)^{(2)}$-Model$.$}

Let us first consider the grading operator $Q= 2d + {{1}\over {2}} h_1$ and $E_{\pm }$ given by 
\br
E_{\pm }= h_1^{(1/2)}+2h_2^{(1/2)} - \left( E_{\alpha _{1}}^{(0)}+E_{-\alpha _{1}}^{(1)} \right) 
\label{e}
\er
which yields  the following  (finite dimensional Lie super algebra)
decomposition
\begin{eqnarray}
\mathcal{K}_{B} &=&\left\{ K_{1}=\lambda _{2}.H,\text{ \ \ }K_{2}=E_{\alpha
_{1}}+E_{-\alpha _{1}}\right\}  \label{sl(2,1) algebra de composition} \\
\mathcal{K}_{F} &=&\left\{ f_{1}=E_{\alpha _{2}}+E_{\alpha _{1}+\alpha _{2}},%
\text{ \ }f_{2}=f_{1}^{\dagger} \text{\ }\right\}  \notag \\
\mathcal{M}_{B} &=&\left\{ M_{1}=h_{1},\text{ \ }M_{2}=E_{\alpha
_{1}}-E_{-\alpha _{1}}\text{\ }\right\}  \notag \\
\mathcal{M}_{F} &=&\left\{ g_{1}=E_{\alpha _{2}}-E_{\alpha _{1}+\alpha _{2},%
\text{ \ \ }}g_{2}=g_{1}^{\dagger}\right\}.  \label{alg21}
\end{eqnarray}
Notice that  the  Cartan subalgebra contribution, $h_1+2h_2\textsl{}$,  in (\ref{e}) 
is crucial in order to generate a nontrivial fermionic  kernel.  
In extending  the Lie super algebra $sl(2,1)$  to a twisted affine structure we assign to each generator 
a loop index $n$ which may be either, integer $n\in Z$ or semi-integer $n\in Z+1/2$.  
In particular this yields  two generators  of grade 1/2 within the Kernel, ${\cal K}_{1/2} = \{ F_1^{(1/2)} = (f_1^{(1/2)}+ f_2^{(1/2)}), \quad F_2^{(1/2)} = f_1^{(1/2)}- f_2^{(1/2)}\} $ and two other generators within the Image, ${\cal M}_{1/2} = \{ G_1^{(1/2)} = g_1^{(1/2)}+ g_2^{(1/2)}, \quad G_2^{(1/2)} = g_1^{(1/2)} - g_2^{(1/2)}\}$.

Since $W_{\pm 1/2}  \in {\mathfrak{g}}_{\pm 1/2}$, they  can be parametrized as
\br
W_{-1/2} = \bar \psi_1 G_1^{(-1/2)} + \bar \psi_2 G_2^{(-1/2)}, \qquad W_{1/2} =  \psi_1 G_1^{(1/2)} +  \psi_2 G_2^{(1/2)}
\label{w}
\er
so that 
\br
Q_{0,+}^{(2)}&=& {{1}\over {2}} [[ E^+, W_{-1/2}], W_{-1/2}] = 4\bar \psi_1 \bar \psi_2 (K_2^{(0)} -K_1^{(0)}), \nonu \\
Q_{0,-}^{(2)}&=& {{1}\over {2}} [[ E^-, W_{1/2}], W_{1/2}] =  4\psi_1  \psi_2 ( K_2^{(0)}-K_1^{(0)})
\label{cond}
\er
In order to satisfy  the subsidiary conditions $Q_{0, \pm}^{(2)} = 0$ we take
\br
\psi_2 = \bar \psi_1 = 0
\nonu
\er
which yields (after renaming $\psi_1 \equiv \psi $ and $\bar \psi_2 \equiv \bar \psi$)
\br
A_{1/2} = -2\psi G_1^{(1/2)}, \quad J_{-1/2} =2 \bar \psi G_2^{(-1/2)}, \qquad B= \exp {\phi M_1^{(0)}}
\label{param}
\er
The action (\ref{SUSY-Toda action}) gives the following Lagrangian
\footnote{%
In 2-dimensions, the Majorana-Weyl basis for the gamma matrices is:%
\begin{equation*}
\gamma ^{0}=\left( 
\begin{array}{cc}
0 & 1 \\ 
-1 & 0%
\end{array}%
\right) \text{, }\gamma ^{1}=\left( 
\begin{array}{cc}
0 & 1 \\ 
1 & 0%
\end{array}%
\right) ,\text{ }\gamma ^{5}=\left( 
\begin{array}{cc}
1 & 0 \\ 
0 & -1%
\end{array}%
\right) .
\end{equation*}%
\par
In the light cone we have $\gamma ^{\pm }=\frac{1}{2}\left( \gamma ^{0}\pm
\gamma ^{1}\right) $ and the projectors $\mathcal{P}^{\pm }=\frac{1}{2}%
\left( 1\pm \gamma ^{5}\right) .$ Define
\par
\begin{equation*}
\Psi =\left( 
\begin{array}{c}
\overline{\psi } \\ 
\psi%
\end{array}%
\right) ,\text{ }\overline{\Psi }=\Psi ^{T}\gamma _{0}.
\end{equation*}}
 density\footnote{%
The Bosonic limit:%
\begin{equation*}
\mathcal{L}=-\frac{k}{2\pi }\left[ \partial _{+}\phi \partial _{-}\phi -2\mu
^{2}\cosh [2\phi ]\right]
\end{equation*}%
gives the Polhmeyer reduction of a Classical Bosonic String in an $%
AdS_{2}\times S^{1}$ background \cite{tseytlin}.}

\begin{eqnarray}
\mathcal{L}_{N=1} &=&-\frac{k}{2\pi }\left[ \partial _{+}\phi \partial
_{-}\phi -\overline{\Psi }\gamma^{\mu} \partial_{\mu} \Psi -V_{N=1}\right]
\label{N=1 super sinh-Gordon} \\
V_{N=1} &=&2\left( \cosh [2\phi ]+\overline{\Psi }\Psi \cosh [\phi ]\right) 
\notag
\end{eqnarray}%
 and the equations of motion (\ref{SUSY-Toda equations}) correspond to  the $N=1$
super sinh-Gordon equations: 
\begin{eqnarray*}
\partial _{+}\partial _{-}\phi &=&2\sinh [2\phi ]+\overline{\Psi }\Psi \sinh
[\phi ] \\
0 &=&\left( \gamma^{\mu} \partial_{\mu} +2\cosh [\phi ]\right) \Psi ,
\end{eqnarray*}%
which are invariant under the supersymmetry transformation (\ref{susy transformations})%
\begin{equation*}
\partial _{1/2}\phi =\epsilon \overline{\psi }\text{ \ \ , \ }\partial _{1/2}%
\overline{\psi }=\epsilon \partial _{x}\phi .
\end{equation*}
{\remark{ Notice that in order to satisfy  the subsidiary conditions (\ref{cond}), 
we have constrained  a pair of Fermi fields, i.e., $\psi_2 = \bar \psi_1 = 0$.  
This, effectively is equivalent in  choosing the subalgebra $sl(2,1)_{\left[ 1\right] }^{(2)}$ (see appendix B)
with only
one  generator within the subspaces ${\mathfrak{g}}_{\pm 1/2}$}.}

Following the same line of thougth
 we now consider the second twisted affine subalgebra $sl(2,1)_{\left[ 2\right] }^{(2)}$ (see appendix B) with 
 parametrization 
\begin{equation*}
B=\exp [\phi M_{2}^{0}],\qquad W_{+1/2}=\psi G_{1}^{1/2},\qquad W_{-1/2}=%
\overline{\psi }G_{2}^{-1/2},
\end{equation*}%
yielding the  Lagrangian density\footnote{%
The Bosonic limit:%
\begin{equation*}
\mathcal{L}=\frac{k}{2\pi }\left[ \partial _{+}\phi \partial _{-}\phi +2\mu
^{2}\cos [2\phi ]\right] ,
\end{equation*}%
gives the Polhmeyer reduction of a Classical Bosonic String in an $%
R_{t}\times S^{2}$ background \cite{tseytlin}.}:%
\begin{eqnarray}
\mathcal{L}_{N=1} &=&\frac{k}{2\pi }\left[ \partial _{+}\phi \partial
_{-}\phi +\overline{\Psi }\gamma^{\mu}  \partial_{\mu} \Psi +V_{N=1}\right]
\label{N=1 super sine-Gordon} \\
V_{N=1} &=&2\left( \cos [2\phi ]+\overline{\Psi }\Psi \cos [\phi ]\right) , 
\notag
\end{eqnarray}%
and  equations of motion given by the $N=1$ super sine-Gordon equations:
\begin{eqnarray*}
\partial _{+}\partial _{-}\phi &=&2\sin [2\phi ]+\overline{\Psi }\Psi \sin
[\phi ] \\
0 &=&\left( \gamma^{\mu} \partial_{\mu} +2\cos [\phi ]\right) \Psi .
\end{eqnarray*}%
invariant under ($\partial _{+}\rightarrow \partial _{x}$) 
\begin{equation*}
\partial _{1/2}\phi =-\epsilon \overline{\psi }\text{ \ \ , \ }\partial
_{1/2}\overline{\psi }=\epsilon \partial _{x}\phi .
\end{equation*}

\begin{remark}
We see that the twisted superalgebra $\hat {sl}(2,1)^{(2)}$ supports the two
models, note also that one is the field analytic continuation of the other
but only at the action level and not in their Lax connections which is where
the models are truly specified \cite{BBT}. We are using a two-valued
spectral parameter $\lambda ^{\pm 1/2}$ due to the gradation used in
contrast to the single-valued Lax pairs usually considered so this $\lambda
^{\pm 1/2}$ plays a role. The Lax connections are different and have cuts,
this opens the question of having a formal Zakharov- Shabat like
construction involving a two-valued spectral parameter.
\end{remark}

\begin{remark}
The untwisted Affine superalgebra $sl(2,1)^{(1)}=\hat{sl}(2,1)$ does not
admit a purely Fermionic simple root system (the longest root being
Bosonic). This means that the usual criteria \cite{olshanetsky}, \cite%
{Hollowood} that only Contragredient Lie superalgebras with a purely
Fermionic simple root system admit supersymmetric Integrable extensions does
not apply when $%
\mathbb{Z}
/2$-gradations are taken into account. The $sl(2,1)^{(2)}$ is an
counter-example of this statement, from it we could construct the $N=1$
super Toda model . We expect this construction help us to have a better
understanding of the intricate relation between supersymmetric Toda models
and Lie superalgebras.
\end{remark}
 
 \subsection{The $N=2$ $\hat {sl}(2,2)^{(2)}$-Model$.$}
 
 Consider the grading operator $Q= d + {{1}\over 2}(h_1+h_3)$ and 
 \be
E_+ = (E_{\a_1}^{(0)} + E_{-\a_1}^{(2)}) + 
(E_{\a_3}^{(2)} + E_{-\a_3}^{(0)}) + I^{(1)} 
\label{esl22a}
\ee
 In the Appendix  B we have found that there are   four
sub-superalgebras of $\hat {sl}(2,2)^{(2)}$ namely  (\ref{Ia})-(\ref{IVa})
which solves $Q_{0\pm }^{2}=0$ . We will write only the ones giving
inequivalent solutions.

For $psl(2,2)_{\left[ 1\right] }^{(2)}$ in (\ref{Ia}), we take the parametrization 
\begin{eqnarray*}
B &=&\exp [\phi _{1}M_{1}^{0}+\phi _{3}M_{3}^{0}] \\
W_{+1/2} &=&\psi _{1}G_{1}^{+1/2}+\psi _{3}G_{3}^{+1/2} \\
W_{-1/2} &=&\overline{\psi }_{2}G_{2}^{-1/2}+\overline{\psi }%
_{4}G_{4}^{-1/2},
\end{eqnarray*}%
to obtain the Lagrangian density%
\br
-{{2\pi }\over {k}} {\mathcal{L}_{N=2}} &=&
\partial _{+}\phi _{1}\partial _{-}\phi _{1}-\partial _{+}\phi _{3}\partial
_{-}\phi _{3} \nonu \\ 
&+& \psi _{1}\partial _{+}\psi _{1}+\overline{\psi }_{2}\partial _{-}\overline{
\psi }_{2}-\psi _{3}\partial _{+}\psi _{3}-\overline{\psi }_{4}\partial _{-}
\overline{\psi }_{4}-V_{N=2}
 \label{N=2 super sinh-Gordon}
 \er 
 where
 \br
V_{N=2} &=&
2\cosh [2\phi _{1}]-2\cosh [2\phi _{3}]-4\left( \psi _{1}\overline{\psi }
_{2}-\psi _{3}\overline{\psi }_{4}\right) \cosh [\phi _{1}]\cosh [\phi _{3}]\nonu 
\\ 
&-& 4\left( \psi _{1}\overline{\psi }_{4}-\psi _{3}\overline{\psi }_{2}\right)
\sinh [\phi _{1}]\sinh [\phi _{3}]
\er
leading to  the $N=2$ super sinh-Gordon model (see for instance \cite{koba-ue}). 

For $psl(2,2)_{\left[ 2\right] }^{(2)}$ in (\ref{IIa}), we have the parametrization 
\begin{eqnarray*}
B &=&\exp [\phi _{1}M_{2}^{0}+\phi _{3}M_{4}^{0}] \\
W_{+1/2} &=&\psi _{1}G_{1}^{+1/2}+\psi _{3}G_{3}^{+1/2} \\
W_{-1/2} &=&\overline{\psi }_{2}G_{2}^{-1/2}+\overline{\psi }%
_{4}G_{4}^{-1/2},
\end{eqnarray*}%
and the Lagrangian density%
\begin{eqnarray}
-\frac{2\pi }{k } {\mathcal{L}_{N=2}} &=&
-\partial _{+}\phi _{1}\partial _{-}\phi _{1}+\partial _{+}\phi _{3}\partial
_{-}\phi _{3} \nonu \\ 
&+&\psi _{1}\partial _{+}\psi _{1}+\overline{\psi }_{2}\partial _{-}\overline{%
\psi }_{2}-\psi _{3}\partial _{+}\psi _{3}-\overline{\psi }_{4}\partial _{-}%
\overline{\psi }_{4}-V_{N=2}%
  \label{N=2 super sine-Gordon} \er
  where
  \br
V_{N=2} &=&
2\cos [2\phi _{1}]-2\cos [2\phi _{3}]-4( \psi _{1}\bar{\psi }_{2}-
\psi _{3}\bar{\psi }_{4}) \cos [\phi _{1}]\cos [\phi _{3}]\nonu 
\\ 
&+&4( \psi _{1}\bar{\psi }_{4}-\psi _{3}\bar{\psi }_{2})
\sin [\phi _{1}]\sin [\phi _{3}])   \notag
\end{eqnarray}%
yields  the $N=2$ super sine-Gordon model.

For $psl(2,2)_{\left[ 3\right] }^{(2)}$ in (\ref{IIIa}), we have the parametrization
\begin{eqnarray*}
B &=&\exp [\phi _{1}M_{1}^{0}+\phi _{3}M_{4}^{0}] \\
W_{+1/2} &=&\psi _{1}G_{1}^{+1/2}+\psi _{3}G_{4}^{+1/2} \\
W_{-1/2} &=&\overline{\psi }_{2}G_{2}^{-1/2}+\overline{\psi }%
_{4}G_{3}^{-1/2},
\end{eqnarray*}%
and the Lagrangian density\footnote{%
This supersymmetric Lagrangian gives the Polhmeyer reduction of a Classical
superstring in an $AdS_{2}\times S^{2}$ background and its Bosonic limit:%
\begin{equation*}
-\frac{2\pi }{k } {\mathcal{L}}= \partial _{+}\phi _{1}\partial _{-}\phi
_{1}+\partial _{+}\phi _{3}\partial _{-}\phi _{3}-\mu ^{2}(2\cosh [2\phi
_{1}]-2\cos [2\phi _{3}])
\end{equation*}%
gives the reduction for a Classical Bosonic String in an $AdS_{2}\times
S^{2} $ background \cite{tseytlin}.}

\begin{eqnarray}
-\frac{2\pi }{k } {\mathcal{L}_{N=2}} &=&
\partial _{+}\phi _{1}\partial _{-}\phi _{1}+\partial _{+}\phi _{3}\partial
_{-}\phi _{3}+ \nonu \\ 
&+&\psi _{1}\partial _{+}\psi _{1}+\overline{\psi }_{2}\partial _{-}\overline{%
\psi }_{2}-\psi _{3}\partial _{+}\psi _{3}-\overline{\psi }_{4}\partial _{-}%
\overline{\psi }_{4}-V_{N=2}%
  \label{N=2 sinh-sine} \er
  where
  \br
V_{N=2} &=&
2\cosh [2\phi _{1}]-2\cos [2\phi _{3}]-4\left( \psi _{1}\overline{\psi }%
_{2}-\psi _{3}\overline{\psi }_{4}\right) \cosh [\phi _{1}]\cos [\phi _{3}]
\nonu \\ 
&+&4\left( \psi _{1}\overline{\psi }_{4}+\psi _{3}\overline{\psi }_{2}\right)
\sinh [\phi _{1}]\sin [\phi _{3}]%
\nonu \end{eqnarray}%
which mixes the $N=1$ sinh-Gordon and sine-Gordon models.

For $psl(2,2)_{\left[ 4\right] }^{(2)}$  we have essentially the same model
as the one obtained with $psl(2,2)_{\left[ 3\right] }^{(2)}$ with the
replacements $\sinh \rightarrow \sin ,$ $\cosh \rightarrow \cos $ and some
sign changes. 

Another interesting example is the pure bosonic composed sinh-sine Gordon model obtained from 
\begin{equation*}
B=\exp [\phi _{1}M_{1}^{0}+\phi _{2}M_{2}^{0}],
\end{equation*}%
and Lagrangian density:
\begin{eqnarray}
\mathcal{L} &=&-\frac{k}{2\pi }\left[ \partial _{+}\phi _{1}\partial
_{-}\phi _{1}-\partial _{+}\phi _{2}\partial _{-}\phi _{2} -V_{\otimes }\right]
\label{new01} \\
V_{\otimes } &=&4 \sinh ^{2}\( \sqrt{ \phi_1^2 - \phi_2^2}\) ,  \notag
\end{eqnarray}%
which  leads to the sinh-Gordon when $\phi_2 \rightarrow 0$ and to the sine-Gordon when $\phi_1 \rightarrow 0$.
\subsection{Solving  $ Q_{0\pm }^{(2)} = 0$}

Let us  assume that the space $\mathcal{M}_{F}$ has the following decomposition: 
\begin{equation}
\mathcal{M}_{F}=\left\{ I_{+}^{N}\dbigoplus I_{-}^{N}\text{ }\mid \text{\ }%
\left[ E,I_{\pm }^{N}\right] \subset \pm I_{\pm }^{N}\text{\ , \ }N=\frac{1}{%
2}\dim \mathcal{M}_{F}=1,2,4...\text{ }\right\} \text{ .}  \label{splitting}
\end{equation}%
This means we have two $adE(\ast )-$ Invariant eigenspaces $I_{\pm }$ of
 dimension $N$ and  eigenvalues  $\pm 1$. Furthermore, we assume that the
following result also holds: 
\begin{equation}
ad^{2}E(\ast )=\Lambda I(\ast )\text{ , \ }\Lambda \in 
\mathbb{R}
^{+}.  \label{equal mass}
\end{equation}

Now we show a situation when $Q_{0 \pm }^{(2)}=0$ happens$.$ We have the following

\begin{proposition}
The conditions $Q_{0\pm }^{(2)}=0$ are satisfied if the adjoint representation
matrix of the constant element $E$ on $\mathcal{M}_{F}$, i.e, $\left[ E,g_{i}%
\right] =\left( AdE\right) _{ij}g_{i}$ decomposes in a block diagonal matrix
of the form $\left[ AdE\right] $ $=diag\left( \mathbb{S},-\mathbb{S}\right) $
, with $\mathbb{S}$ an $N\times N$ symmetric matrix commuting with the
structure constants $\ M$, $N$ and $R$ defined by $\left\{
g_{i}^{+},g_{j}^{+}\right\} =M_{ij}^{a}X_{a}^{0}$ , $\left\{
g_{i}^{-},g_{j}^{-}\right\} =N_{ij}^{b}Y_{b}^{0}$ and $\left\{
g_{i}^{+},g_{j}^{-}\right\} =R_{ij}^{c}(Z_{c}^{0})$, i.e, $\left[ \mathbb{S}%
,M^{a}\right] =\left[ \mathbb{S},N^{b}\right] =\left[ \mathbb{S},R^{c}\right]
=0$ , $i,j=1,...,N$ where $g_{i}^{\pm }$ $\in $ $I_{\pm }^{N}$.
\end{proposition}

\begin{proof}
The proof is made by direct computation. We write a general element $W=\psi
_{i}^{+}g_{i}^{+}+\psi _{i}^{-}g_{i}^{-}$ where $g^{\pm }$ $\mathcal{\in }%
I_{\pm }^{N}$ are such that $\left[ E,g_{i}^{\pm }\right] =\pm \mathbb{S}%
_{ij}g_{i}^{\pm }$. Now we see that $Q_{0}^{2}=\frac{1}{2}\left[ W,\left[ W,E%
\right] \right] $ can be written as 
\begin{equation*}
4Q_{0}=\left( \overrightarrow{\psi }_{+}^{T}\left[ \mathbb{S},M^{a}\right] 
\overrightarrow{\psi }_{+}\right) X_{a}^{0}+\left( \overrightarrow{\psi }%
_{-}^{T}\left[ \mathbb{S},N^{b}\right] \overrightarrow{\psi }_{-}\right)
Y_{b}^{0}+2\left( \overrightarrow{\psi }_{+}^{T}\left[ \mathbb{S},R^{c}%
\right] \overrightarrow{\psi }_{-}\right) Z_{c}^{0}\text{ ,}
\end{equation*}%
from which the statement follows.
\end{proof}

 As an illustration, let us  consider the  $sl(2,1)$ case where $N=1, g_1^+ =g_1 = E_{\a_2} + E_{\a_1+\a_2}$ and  $ g_i^- =g_2 = E_{-\a_2} + E_{-\a_1-\a_2}$.   
 The commutation relations for the odd generators defined in (\ref{alg21}) can be evaluated using the matrix 
 representation given in the appendix B.   We therefore find,
 \br
 \left[ E_+, g_1^+\right] =2g_1^+ , \qquad \left[ E_+, g_1^-\right] = -2g_1^-, 
 \er 
 and $ \mathbb{S}$ is diagonal which commutes with any other matrices.

Another example is  $sl(2,2)$ where $N=2$. The adjoint representation of $E$ is defined by
\br
 \left[ E_+, g_i^+\right] = 2g_i^+, \qquad \left[ E_+, g_i^-\right] = -2g_i^-, \quad i=1,2
 \er 
 where $g_1^+ = g_1+g_4, \quad g_2^+ =g_2+g_3, \quad g_1^- = g_1-g_4, \quad g_2^-= g_2-g_3$ are defined in (\ref{gg}).  
 It therefore follows from the algebra given in the appendix of re. \cite{n2} that again $ \mathbb{S}$ is proportional to the unit matrix 
  and commutes with all other matrices.

\section{$Q_{0, \pm}^{(2)} \neq 0$ and Pure Fermionic Models }
We now discuss  the construction of pure fermionic theories by considering the {\it {maximal Kernel condition}}
 i.e., ${\cal M} _{B} = \emptyset$.  The   bosonic fields in this cases 
 lie in ${\cal K}_{B}$ and may be gauged away by considering the coset $G/{\cal K}_{B}$. 
As a general prototype, consider the super algebra $sl(p,1)$ with homogeneous gradation $Q=d$ and $E^+ = \l_p \cdot H^{(1)}$ (where $\l_p$ is the $p-th$ fundamental weight) such that 
\br
Ker_{E^+} = {\cal K}_{B} = sl(p)\otimes u(1), \qquad {\cal M}_B = \emptyset
\label{1}
\er
and fermionic subspace  generated by $
 \{ E_{\a_p + \cdots + \a_i},  E_{-(\a_p + \cdots + \a_i)},  i=1, 2, \cdots p \}$.
From (\ref{full action}), we now construct the   gauged WZNW  action 
\begin{eqnarray}
& &S[W_{\pm 1/2},B,A_{\pm }]_{G/{\cal K}_{B}^{0}} =S_{WZNW}[B] + \frac{k}{2\pi }\int \left\langle \Phi^{-1} E_+ \Phi B \Psi E_- \Psi^{-1} B^{-1}  \right\rangle \nonu \\
&-&\frac{k}{4\pi }\int
\left\langle D_{-}W_{-1/2}\left[ W_{-1/2},E_{+}\right] -D_{+}W_{+1/2}\left[
W_{+1/2},E_{-}\right] \right\rangle +  
\nonu \\
&-&\frac{k}{2\pi }\int \left\langle  A_{-}\left( \partial
_{+}BB^{-1}\right) +A_{+}\left( B^{-1}\partial _{-}B\right) +
A_{-}BA_{+}B^{-1}+A_{-}^{0}A_{+}^{0}\right\rangle , \nonu \\
 \label{g-action}
\er
where 
\br
D_{-}W_{-1/2} &=&\partial _{-}W_{-1/2}+ \left[ A_{-},W_{-1/2}\right] 
\notag \\
D_{+}W_{+1/2} &=&\partial _{+}W_{+1/2}-\left[ A_{+},W_{+1/2}\right] \nonu
\er
$A_{\pm }=A_{\pm }^{0}+N_{\pm }^{0}\in $ ${\cal{K}}_{B}^{0}$, where $A_{\pm }^{0}$ lies in the Cartan subalgebra of $sl(p)\otimes u(1)$ and  $N_{\pm }^{0}$ in its orthogonal  complement, are the auxiliary gauge
fields and $D_{\pm }$ the covariant derivatives.  The action (\ref{g-action}) 
 is invariant under the following transformation
\begin{eqnarray}
A_{-}^{\prime 0} &=&A_{-}^{0}- \gamma _{0}^{-1}\partial _{-}\gamma _{0}
\label{gauge transformations 2} \\
A_{+}^{\prime 0} &=&A_{+}^{0}-\gamma _{0}^{ -1}\partial _{+}\gamma
_{0}^{ }  \notag \\
A_{-}^{\prime } &=&\Gamma _{-}A_{-}\Gamma _{-}^{-1}- \partial _{-}\Gamma
_{-}\Gamma _{-}^{-1}  \notag \\
A_{+}^{\prime } &=&\Gamma _{+}^{-1}A_{+}\Gamma _{+}-\Gamma _{+}^{-1}\partial
_{+}\Gamma _{+}  \notag 
\end{eqnarray}
and 
\begin{eqnarray}
B^{\prime } &=&\Gamma _{-}B\Gamma _{+}  \label{gauge transformations 1} \\
W_{-1/2}^{\prime } &=&\Gamma _{-}W_{-1/2}\Gamma _{-}^{-1}  \notag \\
W_{+1/2}^{\prime } &=&\Gamma _{+}^{-1}W_{+1/2}\Gamma _{+},  \label{invariance}
\end{eqnarray}%
where $\Gamma _{-}=\gamma _{0}\gamma _{-}$ and $\Gamma _{+}=\gamma
_{+}\gamma _{0}\ $are both in ${\cal K}_{B}$.
Here $\g_0, \g_{\pm}$ are exponentials of Cartan subalgebra and positive/negative step operators of $sl(p)\otimes u(1)$ respectively.
The  two terms involving covariant derivatives in (\ref{g-action})  are clearly invariant.  
The  last term was  introduced  in \cite{isospin} within the  coset structure $sl(3)/sl(2)\otimes u(1)$ and   
its invarince under the ${\mathfrak{g}}_{0} = sl(2)\otimes u(1)$ was  shown to lead to Isospin conservation laws.   

The invariance of the action (\ref{g-action}) under (\ref{invariance}) allows us to choose a gauge in which $B^{\prime} =I$.
Under such circunstances, we find
\begin{eqnarray}
 & &S[W_{\pm 1/2}^{\pr},B=I,A_{\pm }^{\pr}]_{G/K_{B}^{0}} = \frac{k}{2\pi }\int \left\langle \Phi^{-1} E_+ \Phi  \Psi E_- \Psi^{-1}   \right\rangle \nonu \\ &-&\frac{k}{4\pi }\int
\left\langle \pa_-W_{-1/2}^{\pr}\left[ W_{-1/2}^{\pr},E_{+}\right] -\pa_+W_{+1/2}^{\pr}\left[
W_{+1/2}^{\pr},E_{-}\right] \right\rangle +  
 \nonu \\
 &-&\frac{k}{2\pi }\int
\left\langle  A_-^{\pr} Q_{0+}^{(2)}+   A_+^{\pr} Q_{0-}^{(2)}  + A_{-}^{\pr }A_{+}^{\pr }  +A_{-}^{\pr 0}A_{+}^{\pr 0}   \right\rangle \nonu \\ 
 \label{g-action1}
\end{eqnarray}%
Since the action is quadratic in the auxiliary fields and $\left\langle A_{\pm }^{ 0} N_{\pm }^{0} \right\rangle =0 $ we can use Gaussian integration to obtain the effective action
\br
S[W_{\pm 1/2}]_{effective} &=& \frac{k}{2\pi }\int \left\langle \Phi^{-1} E_+ \Phi  \Psi E_- \Psi^{-1}  +{\frac{3}{2}}Q_{0+}^{(2)} Q_{0-}^{(2)} \right\rangle \nonu \\ &-&\frac{k}{4\pi }\int
\left\langle \pa_-W_{-1/2}\left[ W_{-1/2},E_{+}\right] -\pa_+W_{+1/2}\left[
W_{+1/2},E_{-}\right] \right\rangle \nonu \\
 \label{g-action2}
\end{eqnarray}%

As an illustration  consider the simplest case where ${\mathfrak{g}} = sl(2,1)$.
Let us parametrize
\br
W_{1/2} &=&\psi _{1}E_{\a_2}^{(1/2)} +\psi _{2}E_{\a_1+\a_2}^{(1/2)} +\psi _{3}E_{-\a_2}^{(1/2)} +\psi _{4}E_{-\a_1-\a_2}^{(1/2)}
\nonu \\
W_{-1/2} &=&\bar \psi _{1}E_{\a_2}^{(-1/2)} +\bar \psi _{2}E_{\a_1+\a_2}^{(-1/2)} +\bar \psi _{3}E_{-\a_2}^{(-1/2)} +\bar \psi _{4}E_{-\a_1-\a_2}^{(-1/2)}
 \label{w}
\er
from where we  evaluate
\br
Q_{0-}^{(2)} &=&\left( \psi _{2}\psi _{4}\right) h_{1}+\left( \psi _{1}\psi
_{3}+\psi _{2}\psi _{4}\right) h_{2}+\left( \psi _{2}\psi _{3}\right)
E_{\a_1}+\left( \psi _{1}\psi _{4}\right) E_{-\a_1}, \nonu \\
Q_{(1/2)-}^{(3)} &=&\frac{1}{3}\left[ W_{1/2},Q_{0-}^{(2)}\right] 
=\frac{2}{3}%
\left( \psi _{1}\psi _{2}\psi _{4}\right) E_{\a_2}^{(1/2)}\nonu \\
&-&\frac{2}{3}\left( \psi
_{1}\psi _{2}\psi _{3}\right) E_{\a_1+\a_2}^{(1/2)}
+\frac{2}{3}\left( \psi _{2}\psi _{3}\psi
_{4}\right) E_{-\a_2}^{(1/2)}, \nonu \\
Q_{(1)-}^{(4)} &=&\frac{1}{4}\left[ W_{1/2},Q_{(1/2)-}^{(3)}\right] =\frac{1}{6}%
\left( \psi _{1}\psi _{2}\psi _{3}\psi _{4}\right) \left(
h_{1}^{(1)}+3h_{2}^{(1)}\right)\label{q4}  \er
and similar for  $Q_{0+}^{(2)}, Q_{(-1/2) +}^{(3)}$ and $Q_{(-1) +}^{(4)}$ in terms    $\bar \psi_i $ fields.
The potential term in the Lagrangian  decomposes according to the number of fermions  involving the following individual contributions,
\br
\left\langle Q_{0+}^{(2)}, Q_{0-}^{(2)}\right\rangle  &=&-\left( \overline{\psi }%
_{2}\overline{\psi }_{4}\right) \left( \psi _{1}\psi _{3}\right) -\left( 
\overline{\psi }_{1}\overline{\psi }_{3}\right) \left( \psi _{2}\psi
_{4}\right) \nonu \\
&+&\left( \overline{\psi }_{2}\overline{\psi }_{3}\right) \left(
\psi _{1}\psi _{4}\right) +\left( \overline{\psi }_{1}\overline{\psi }%
_{4}\right) \left( \psi _{2}\psi _{3}\right) \nonu  \\
\left\langle Q_{(-1/2)+}^{(3)}Q_{(1/2)-}^{(3)}\right\rangle  &=&-\frac{4}{9}%
\left( \left( \overline{\psi }_{1}\overline{\psi }_{2}\overline{\psi }%
_{4}\right) \left( \psi _{2}\psi _{3}\psi _{4}\right) -\left( \overline{\psi 
}_{2}\overline{\psi }_{3}\overline{\psi }_{4}\right) \left( \psi _{1}\psi
_{2}\psi _{4}\right) \right)  \nonu \\
\left\langle Q_{(-1)+}^{(4)}Q_{(1)-}^{(4)}\right\rangle  &=&-\frac{1}{9}\left( 
\overline{\psi }_{1}\overline{\psi }_{2}\overline{\psi }_{3}\overline{\psi }%
_{4}\right) \left( \psi _{1}\psi _{2}\psi _{3}\psi _{4}\right) 
\label{qqq}
\er
The Lagrangian density with normalized  coupling constants {\footnote{ Coupling constants $g$ and $\mu$ may be introduced by $W_{\pm 1/2} \rightarrow gW_{\pm 1/2}$ and $E_{\pm} \rightarrow \mu E_{\pm}$}} then becomes
\br
-{{2\pi }\over {k}} L_{p=2} &=& \psi_1 \pa_+ \psi_3 -  \bar \psi_1 \pa_- \bar \psi_3
+  \psi_2 \pa_+ \psi_4 -  \bar \psi_2 \pa_- \bar \psi_4 \nonu \\
&-&
\bar \psi_1   \psi_3 +  \bar \psi_3   \psi_1 - \bar \psi_2 \psi_4 +  \bar \psi_4   \psi_2
 +
\bar \psi_2 \bar \psi_4 \psi_1 \psi_3 \nonu \\
&+& \bar \psi_1 \bar \psi_3 \psi_2 \psi_4
- \bar \psi_2 \bar \psi_3 \psi_1 \psi_4 - \bar \psi_1 \bar \psi_4 \psi_2 \psi_3\nonu \\
&-& {{4}\over {9}} \bar \psi_1\bar \psi_2\bar \psi_4 \psi_2\psi_3\psi_4 +{{4}\over {9}} \bar \psi_2\bar \psi_3\bar \psi_4 \psi_1\psi_2\psi_4 \nonu \\
&-& {{1}\over {9}} \bar \psi_1\bar \psi_2\bar \psi_3\bar \psi_4 \psi_1\psi_2\psi_3\psi_4 
\label{8}
\er
which can be put in the Dirac form
\begin{eqnarray*}
L &=&\bar{\Psi}_D \left( i\gamma^{\mu}  \partial_{\mu} -1\right) \Psi_D  
+\bar{\Phi}_D \left( i\gamma^{\mu}  \partial_{\mu} -1\right) \Phi_D  \nonu \\
&+& \frac{5}{4} \left( \bar{\Psi}_D  \gamma^{\mu }\Psi_D  \right) \left( \bar{\Phi_D }\gamma _{\mu }\Phi_D \right)
+\frac{5}{4} \left( \bar{\Psi}_D  \Psi_D  \right) \left( \bar{\Phi}_D \Phi_D \right) \nonu \\
& -&\frac{5}{4} \left( \bar{\Psi}_D  \gamma ^{5}\Psi_D  \right) \left( \bar{\Phi}_D 
\gamma ^{5}\Phi_D \right)  
-\frac{1}{9}\left( \bar{\Phi}_D \gamma ^{\mu }\Phi_D \right) \left( 
\bar{\Phi}_D \gamma _{\mu }\Phi_D \right) \left( \bar{\Psi}_D  \Psi_D 
\right) \nonu \\
&+&\frac{1}{144}\left( \bar{\Phi}_D \gamma ^{\mu }\Phi_D \right)
\left( \bar{\Phi}_D \gamma _{\mu }\Phi_D \right) \left( \bar{\Psi}_D  
\gamma ^{\nu }\Psi_D  \right) \left( \bar{\Psi}_D  \gamma _{\nu }\Psi_D  \right) 
\label{basic model}
\end{eqnarray*}
where  the complex Dirac spinor components are 
 $\Psi_{D 1}=i\psi
_{3}$ , $\Psi_{D 2}=-{\bar \psi }_{3}$ , $\Psi_{D 1}^{+}=-\psi _{1}$, $
\Psi_{D 2}^{+}=-i\bar{\psi }_{1}$ and $\Phi_{D 1}=i\psi _{4}$ , $\Phi_{D 2}=-\bar{\psi }_{4}$ , $\Phi_{D 1}^{+}=-\psi _{2}$ , $\Phi_{D 2}^{+}=-i
\bar{\psi }_{2}$.

\subsection{Zero Curvature Representation and Integrability Conditions}

We now discuss the integrability conditions of the class of models constructed so far.
Inspired by the Lax form when $Q_{0 \pm}^{(2)} =0$ in (\ref{lax}), propose the following structure
\br
\tilde{L}_{+1} = \pa_+ + E_+ +A_{1/2} + X Q_{0+}^{(2)} ,\qquad 
\tilde{L}_{-1} = \pa_- - E_- - J_{-1/2} +Y Q_{0-}^{(2)}
\label{lax2}
\er 
where $A_{1/2}, J_{-1/2}, Q_{0 \pm }^{(2)}$ are defined in terms of $W_{\pm 1/2}$ as in (\ref{fields conection}) and (\ref{cond}), 
$X$ and $Y$ are parameters to be ajusted.  The zero curvature condition (\ref{zcc}) yields the following equations,
\br
\pa_- A_{1/2} &=& \lb E_+, J_{-1/2} \rb + Y\lb A_{1/2}, Q_{0 -}^{(2)}\rb, \label{compat1}\\
\pa_+ J_{-1/2} &=& -\lb A_{1/2}, E_- \rb - X\lb  Q_{0 +}^{(2)}, J_{-1/2} \rb, \label{compat2}\\
X\pa_+ Q_{0-}^{(2)} + Y\pa_-  Q_{0+}^{(2)} &=& -\lb A_{1/2}, J_{-1/2}\rb + XY\lb Q_{0+}^{(2)}, Q_{0-}^{(2)}\rb 
\label{compatibil}
\er
Consider now 
the equations of motion derived  from the Lagrangian (\ref{g-action2}) when $ Q_{(\mp1/2) \pm }^{(3)} = Q_{(\mp 1)\pm }^{(4)} =0$,
\br
\pa_- A_{1/2} &=& -\lb E_+, J_{-1/2} \rb + \lb A_{1/2}, Q_{0 -}^{(2)}\rb, \nonu \\
\pa_+ J_{-1/2} &=& -\lb A_{1/2}, E_- \rb + \lb  Q_{0 +}^{(2)}, J_{-1/2} \rb, \nonu \\
\label{qq}
\er
from where we derive
\br
\pa_- W_{-1/2} = J_{-1/2} + \lb W_{-1/2}, Q_{0 -}^{(2)}\rb, \qquad 
\pa_+ W_{1/2} = A_{1/2} - \lb W_{1/2}, Q_{0 +}^{(2)}\rb
\label{ww}
\er
Using the the definition of  $Q_{0 \pm}^{(2)}$, we find  after employing eqns. (\ref{qq}) and (\ref{ww}) 
and Jacobi identities,
\br
Y\pa_+ Q_{0-}^{(2)} + X\pa_-  Q_{0+}^{(2)} &=& (X+Y) \(\lb A_{1/2}, J_{-1/2}\rb +2 \lb Q_{0 +}^{(2)},Q_{0 -}^{(2)}\rb\)
\label{ccon}
\er
where we have used  the fact that, for ${\mathfrak{g}}= sl(p,1)$, $\lb J_{-1/2}, W_{-1/2}\rb \in {\cal K}$ and $\lb A_{1/2}, W_{1/2}\rb \in {\cal K}$ and therefore 
$\lb E_{+}, \lb J_{-1/2}, W_{-1/2}\rb \rb = \lb E_{-}, \lb A_{1/2}, W_{1/2}\rb \rb =0$.
Comparing (\ref{ccon}) with the l.h.s. of (\ref{compatibil}) we find compatibility between zero curvature representation 
and equations of motion if $X+Y = 1$  and 
\br
  \lb Q_{0+}^{(2)}, Q_{0-}^{(2)}\rb =0.
\label{integrr}
\label{condit}
\er
Notice that the Lax (\ref{lax2}), $\tilde{L}_{\pm1}$  coincide with $L_{\pm 1}$ of eqns. (\ref{a}) and (\ref{b}) when $B=I, A_0 =0$ for $X=1, \;\; Y=0$.
Other solutions of $X+Y=1$ correspond to different models related by non local  gauge transformations 
involving $Q_{0 \pm}^{(2)} \in {\cal K}$.   
\subsection{Examples}

We now consider some explicit examples  where $Q_{(\mp1/2) \pm }^{(3)} = Q_{(\mp 1)\pm }^{(4)} =0  $  constructed by imposing  the integrability conditions (\ref{integrr})
to the basic model (\ref{basic model}).  In order to fullfil the integrability conditions (\ref{integrr})
let us consider the following different cases by taking for instance  $X=Y = 1/2$:
\subsubsection{  Constrained Bukhvostov-Lipatov Model}
Take the constraint
\br
\psi_2  \psi_3 = \psi_1  \psi_4 = 0, \qquad \bar \psi_2  \bar \psi_3 = \bar \psi_1 \bar \psi_4=0
\label{B-L}
\er
Under such conditions we find from eqns. (\ref{q4}) and (\ref{qqq}), 
\br
Q^{(2)}_{0-} &=&  \psi _{2}\psi _{4} h_{1}+( \psi _{1}\psi
_{3}+\psi _{2}\psi _{4}) h_{2}\nonu \\
Q^{(2)}_{0+} &=&  \bar \psi _{2}\bar \psi _{4} h_{1}+(\bar  \psi _{1}\bar \psi
_{3}+\bar \psi _{2}\bar \psi _{4}) h_{2}
\label{bl1}
\er
and 
\br
\left\langle Q_{0+}^{(2)}, Q_{0-}^{(2)}\right\rangle  &=&-\bar{\psi }
_{2}\bar{\psi }_{4}  \psi _{1}\psi _{3} -
\bar{\psi }_{1}\bar{\psi }_{3} \psi _{2}\psi_{4}.
\er
The Lagrangian density becomes \cite{lipatov},
\br
L &=&\bar{\Psi}_D ( i\gamma^{\mu}  \partial_{\mu} -1) \Psi_D  +\bar{
\Phi}_D ( i\gamma^{\mu}  \partial_{\mu} -1) \Phi_D  \nonu \\
&&+\frac{5}{4}  ( \bar{\Psi}_D  \gamma
^{\mu }\Psi_D  ) ( \bar{\Phi}_D \gamma _{\mu }\Phi_D )
\label{bl-lagr}
\er
with fields satisfying  constraint (\ref{B-L}).

\subsubsection{  Thirring Model}
Take now a solution of  (\ref{B-L}), namely,
\br
\psi_1 = - \psi_4, \quad \psi_2 = - \psi_3, \qquad \bar \psi_1 = \bar \psi_4, \quad \bar \psi_2 = \bar \psi_3
\label{thirr}
\er
\br
Q^{(2)}_{0-} =  \psi _{2}\psi _{4} \;\;h_{1}, \qquad 
Q^{(2)}_{0+} =  \bar \psi _{2}\bar \psi _{4} \;\;h_{1}
\label{thirr1}
\er
yielding the Lagrangian density
\br
L =\bar{\Psi}_D \left( i\gamma^{\mu}  \partial_{\mu} -1\right) \Psi_D  
+  (\bar{\Psi}_D  \gamma^{\mu }\Psi_D)( \bar{\Psi}_D \gamma _{\mu }\Psi_D) 
\label{thirr-lagr}
\er
which correspond to the Thirring model.
\subsubsection{Pseudo scalar, massless Gross-Neveu model}
Consider the constraint 
\br
\bar \psi_2 = \bar \psi_4, \quad \bar \psi_1 = \bar \psi_3, \quad
 \psi_2 =  \psi_4, \quad  \psi_1 =  \psi_3
 \label{g-n}
 \er
 giving
 \br
Q^{(2)}_{0-} =  -\psi _{1}\psi _{2} \;\;h_{1}, \qquad 
Q^{(2)}_{0+} =  -\bar \psi _{1}\bar \psi _{2} \;\;h_{1}
\label{gn1}
\er
and therefore yielding the Gross-Neveu model,
\br
L &=&\bar{\Psi}_D \left( i\gamma^{\mu}  \partial_{\mu} \right) \Psi_D  +\bar{
\Phi}_D \left( i\gamma^{\mu}  \partial_{\mu} \right) \Phi_D  
-   \left( \bar{\Psi}_D  \gamma^5 \Psi_D  \right) \left( \bar{\Phi}_D \gamma ^{5 }\Phi_D \right)
\label{gn-lagr}
\er

\subsubsection{Scalar, Massive Gross-Neveu model}
Consider
\br
\bar \psi_2 = \bar \psi_4, \quad \bar \psi_1 = \bar \psi_3, \quad
 \psi_2 =  -\psi_4, \quad  \psi_1 =  -\psi_3
 \label{g-nm}
 \er
 with 
 \br
Q^{(2)}_{0-} =  \psi _{1}\psi _{2} \;\;(E_{\a_1} - E_{-\a_1}), \qquad 
Q^{(2)}_{0+} =  \bar \psi _{1}\bar \psi _{2} \;\;(E_{\a_1} - E_{-\a_1})
\label{gnm1}
\er
yielding the Gross-Neveu model,
\br
L &=&\bar{\Psi}_D \left( i\gamma^{\mu}  \partial_{\mu} -1\right) \Psi_D  +\bar{
\Phi}_D \left( i\gamma^{\mu}  \partial_{\mu} -1\right) \Phi_D  
-   \left( \bar{\Psi}_D  \Psi_D  \right) \left( \bar{\Phi}_D \Phi_D \right)
\label{gnm-lagr}
\er

\section{Concluding remarks}

Based on the gauged WZNW model, 
we have established a general framework for constructing  systematicaly the action 
for a class of $N=1,2$ supersymmetric relativistic integrable models of sinh(sine)-Gordon type. 
It is important to stress that the field content of the theory is established by 
the group theoretic structure of a coset $G/{\cal K}$ and the latter by the 
decomposition of an twisted affine super Kac-Moody algebra.
Whithin this context, it would be interesting to uncover the algebraic structure 
of  affine supersymmetric integrable models with higher 
supersymmetries, $N>2$ as well as   higher Toda models.   

Moreover,  the higher grading generalization of the Toda systems proposed by Gervais and 
Saveliev and its connection to matter fields arises naturally from the coset  and the gauged WZNW 
structures.

Another important achievement of our formalism is the construction of 
pure fermionic theories by considering the  coset $sl(p,1)/sl(p)\otimes U(1)$ 
where all bosonic fields lie in the {\it maximal kernel} subalgebra ${\cal K} = sl(p)\otimes U(1)\in sl(p,1)$. 
General integrability conditions were discussed and explicit examples for $p=2$ were constructed.
The generalization  for $p>2$ would be  interesting and would lead to other fermionic integrable theories.

{\bf Acknowledgments} 
JFG and AHZ aknowledges CNPq for partial support.
 D.M.S thanks FAPESP for  financial support and the
IPhT/Saclay for its hospitality and the very stimulating scientific
atmosphere where part of this work was done. D.M.S is gratefull to A.
Tseytlin for comments and suggestions. 
We thank H. Aratyn, A.R. Aguirre and L.H. Ymai  for discussions.

\section{Appendix A: Invariance of the Toda action under the SUSY Flow.}

Here we prove that the 
 action (\ref{SUSY-Toda action}) is invariant under the $1/2$-grade 
flow  (\ref{SUSY transformations}) generated by the element $D^{1/2} \in {\cal K}$, namely 
\begin{eqnarray}
\partial _{1/2}J_{-1/2} &=&\left[ E_{-},B^{-1}D^{1/2}B\right]  \notag \\
\partial _{-}\left( \partial _{1/2}BB^{-1}\right) &=&\left[
D^{1/2},BJ_{-1/2}B^{-1}\right]  \nonu  \\
\left[ E_{+},\partial _{1/2}BB^{-1}\right] &=&\left[ A_{1/2},D^{1/2}\right]
\label{3} \\
\partial _{1/2}A_{1/2} &=&\left[ \partial _{1/2}BB^{-1},A_{1/2}\right] -%
\left[ \partial _{+}BB^{-1},D^{1/2}\right],   \label{2}
\end{eqnarray}
 i.e., $ \pa_{1/2} S_{Toda}=0.$
We will assume that (\ref{equal mass}) holds. This allows to rewrite (\ref{3}%
) as $\partial _{1/2}BB^{-1}=\left[ D^{1/2},W_{-1/2}\right] $ which is $%
\Lambda $-independent.  The relevant transformations are then 
given by 
\begin{eqnarray}
\partial _{1/2}J_{-1/2} &=&\left[ E_{-},B^{-1}D^{1/2}B\right] \nonu \\
\partial _{1/2}BB^{-1} &=&\left[ D^{1/2},W_{-1/2}\right]  \nonu \\
\partial _{1/2}A_{1/2} &=&\left[ \left[ D^{1/2},W_{-1/2}\right] ,A_{1/2}
\right] -\left[ \partial _{+}BB^{-1},D^{1/2}\right].  \label{fundamental SUSY trans} 
\end{eqnarray}

An arbitrary variation of the action (\ref{SUSY-Toda action}) is given by%
\begin{eqnarray*}
\frac{2\pi }{k}\delta S_{Toda} &=&\dint \left\langle \left( \delta
BB^{-1}\right) \left\{ \partial _{-}\left( \partial _{+}BB^{-1}\right) -%
\left[ E_{+},BE_{-}B^{-1}\right] -\left[ A_{1/2},BJ_{-1/2}B^{-1}\right]
\right\} \right\rangle + \\
&&+\left\langle \delta J_{-1/2}\left\{ \partial
_{+}W_{+1/2}-B^{-1}A_{1/2}B\right\} \right\rangle +\left\langle \delta
A_{1/2}\left\{ -\partial _{-}W_{-1/2}+BJ_{-1/2}B^{-1}\right\} \right\rangle ,
\end{eqnarray*}%
taking $\delta \rightarrow \partial _{1/2}$ and using (\ref{fundamental SUSY
trans}) we have%
\begin{eqnarray*}
\frac{2\pi }{k}\partial _{1/2}S_{Toda} &=&\dint \left\langle \left[
D^{1/2},W_{-1/2}\right] \left\{ \partial _{-}\left( \partial
_{+}BB^{-1}\right) -\left[ E_{+},BE_{-}B^{-1}\right] -\left[
A_{1/2},BJ_{-1/2}B^{-1}\right] \right\} \right\rangle + \\
&&+\left\langle \left[ E_{-},B^{-1}D^{1/2}B\right] \left\{ \partial
_{+}W_{+1/2}-B^{-1}A_{1/2}B\right\} \right\rangle + \\
&&+\left\langle \left\{ \left[ \left[ D^{1/2},W_{-1/2}\right] ,A_{1/2}\right]
-\left[ \partial _{+}BB^{-1},D^{1/2}\right] \right\} \left\{ -\partial
_{-}W_{-1/2}+BJ_{-1/2}B^{-1}\right\} \right\rangle .
\end{eqnarray*}
Analising term by term, we find  using the identity $\left\langle a\left[ b,c\right] \right\rangle
=\left\langle \left[ a,b\right] c\right\rangle $ we obtain a total derivative%
\begin{eqnarray*}
&&\left\langle \left[ D^{1/2},W_{-1/2}\right] \partial _{-}\left( \partial
_{+}BB^{-1}\right) +\left[ \partial _{+}BB^{-1},D^{1/2}\right] \partial
_{-}W_{-1/2}\right\rangle \\
&=&\left\langle \partial _{-}\left( \partial _{+}BB^{-1}\right) \left[
D^{1/2},W_{-1/2}\right] \right\rangle +\left\langle \left( \partial
_{+}BB^{-1}\right) \left[ D^{1/2},\partial _{-}W_{-1/2}\right] \right\rangle
\\
&=&\partial _{-}\left\langle \left( \partial _{+}BB^{-1}\right) \left[
D^{1/2},W_{-1/2}\right] \right\rangle \text{.}
\end{eqnarray*}

The term
\begin{equation*}
-\left\langle \left[ \left[ D^{1/2},W_{-1/2}\right] ,A_{1/2}\right] \partial
_{-}W_{-1/2}\right\rangle =-\left\langle \left[ D^{1/2},W_{-1/2}\right] %
\left[ A_{1/2},\partial _{-}W_{-1/2}\right] \right\rangle =0,
\end{equation*}%
since $\left[ A_{1/2},\partial _{-}W_{-1/2}%
\right] \in \mathcal{K}_{B}$ ,$\left[ D^{1/2},W_{-1/2}\right] \in \mathcal{M}%
_{B}$ and  the  orthogonality $\left\langle \mathcal{K}_{B}\mathcal{M}_{B}\right\rangle =0$ 

 The term%
\begin{equation*}
-\left\langle \left[ D^{1/2},W_{-1/2}\right] \left[ A_{1/2},BJ_{-1/2}B^{-1}%
\right] \right\rangle +\left\langle \left[ \left[ D^{1/2},W_{-1/2}\right]
,A_{1/2}\right] BJ_{-1/2}B^{-1}\right\rangle =0
\end{equation*}%
is also zero because of the identity $\left\langle a\left[ b,c\right]
\right\rangle =\left\langle \left[ a,b\right] c\right\rangle .$

For the term%
\begin{eqnarray*}
&&\left\langle -\left[ D^{1/2},W_{-1/2}\right] \left[ E_{+},BE_{-}B^{-1}%
\right] -\left[ E_{-},B^{-1}D^{1/2}B\right] B^{-1}A_{1/2}B\right\rangle \\
&=&-2\left\langle \left[ D^{1/2},W_{-1/2}\right] \left[ E_{+},BE_{-}B^{-1}%
\right] \right\rangle =-2\left\langle E_{-}B^{-1}\left[ D^{1/2},A_{1/2}%
\right] B\right\rangle
\end{eqnarray*}%
we use the Jacobi identity $\left[ E_{+},\left[ BE_{-}B^{-1},D^{1/2}\right] %
\right] =-\left[ D^{1/2},\left[ E_{+},BE_{-}B^{-1}\right] \right] $. We also
have%
\begin{eqnarray*}
&&\left\langle \left[ E_{-},B^{-1}D^{1/2}B\right] \partial _{+}W_{+1/2}-%
\left[ \partial _{+}BB^{-1},D^{1/2}\right] BJ_{-1/2}B^{-1}\right\rangle \\
&=&-\partial _{+}\left\langle \left( B^{-1}D^{1/2}B\right)
J_{-1/2}\right\rangle +2\left\langle BJ_{-1/2}B^{-1}\left[ D^{1/2},\partial
_{+}BB^{-1}\right] \right\rangle .
\end{eqnarray*}

Then, we have that the remaining sum%
\begin{eqnarray*}
&&-2\left\langle \left[ D^{1/2},A_{1/2}\right] BE_{-}B^{-1}\right\rangle
+2\left\langle \left[ D^{1/2},\partial _{+}BB^{-1}\right] BJ_{-1/2}B^{-1}%
\right\rangle \\
&=&2\left\langle D^{1/2}B\partial _{+}J_{-1/2}B^{-1}\right\rangle
+2\left\langle D^{1/2}\left[ \partial _{+}BB^{-1},BJ_{-1/2}B^{-1}\right]
\right\rangle \\
&=&2\partial _{+}\left\langle D^{1/2}\left( BJ_{-1/2}B^{-1}\right)
\right\rangle ,
\end{eqnarray*}%
is total derivative because of the equations of motion for the field $%
J_{-1/2}$ and the identity $\left\langle a\left[ b,c\right] \right\rangle
=\left\langle \left[ a,b\right] c\right\rangle $. The result then follows.

\section{Appendix B: Usefull quantities.}

Here we present the the structure  of some of the Lie
superalgebras used above. The  simple root system can be constructed 
from the following basis $\left( e_{i},\zeta_{k}\right) $ of a real pseudo-euclidean $\left( m+n\right) -$%
dimensional space:%
\begin{equation*}
e_{i}.e_{j}=\delta _{ij},\qquad \zeta_{k}.\zeta_{l}=-\delta _{kl},\qquad
e_{i}.\zeta_{k}=0\text{ ,}
\end{equation*}%

The non-degenerate Cartan matrix is given by $K_{ij}=\left( \alpha
_{i},\alpha _{j}\right) $. 
The commutation relations are 
\begin{eqnarray*}
\left[ h_{i},E_{\pm \alpha _{j}}\right] &=&\pm K_{ij}E_{\pm \alpha _{j}} \\
\left[ E_{\alpha _{i}},E_{-\alpha _{j}}\right] &=&\delta _{ij}h_{i} \\
\left[ h_{i},h_{j}\right] &=&0\text{ .}
\end{eqnarray*}

We will write explicitly the representation matrices only for $sl(2,1)$. For
the other ones a similar obvious construction holds.

\subsection{The sl(2,1) superalgebra.}

This algebra can be represented by $3\times 3$  matrices with 4 Bosonic
elements:

\begin{eqnarray*}
h_{1} &=&\left( 
\begin{array}{ccc}
1 & 0 & 0 \\ 
0 & -1 & 0 \\ 
0 & 0 & 0%
\end{array}%
\right) \text{ \ }h_{2}=\left( 
\begin{array}{ccc}
0 & 0 & 0 \\ 
0 & 1 & 0 \\ 
0 & 0 & 1%
\end{array}%
\right) \text{ \ } \\
E_{\alpha _{1}} &=&\left( 
\begin{array}{ccc}
0 & 1 & 0 \\ 
0 & 0 & 0 \\ 
0 & 0 & 0%
\end{array}%
\right) \text{ \ }E_{-\alpha _{1}}=\left( 
\begin{array}{ccc}
0 & 0 & 0 \\ 
1 & 0 & 0 \\ 
0 & 0 & 0%
\end{array}%
\right) \text{,\ }
\end{eqnarray*}%
and 4 the Fermionic ones: 
\begin{eqnarray*}
E_{\alpha _{2}} &=&\left( 
\begin{array}{ccc}
0 & 0 & 0 \\ 
0 & 0 & 1 \\ 
0 & 0 & 0%
\end{array}%
\right) \text{ \ }E_{-\alpha _{2}}=\left( 
\begin{array}{ccc}
0 & 0 & 0 \\ 
0 & 0 & 0 \\ 
0 & 1 & 0%
\end{array}%
\right) \text{ \ \ } \\
E_{\alpha _{1}+\alpha _{2}} &=&\left( 
\begin{array}{ccc}
0 & 0 & 1 \\ 
0 & 0 & 0 \\ 
0 & 0 & 0%
\end{array}%
\right) \text{ \ }E_{-\alpha _{1}-\alpha _{2}}=\left( 
\begin{array}{ccc}
0 & 0 & 0 \\ 
0 & 0 & 0 \\ 
1 & 0 & 0%
\end{array}%
\right) \text{ \ \ \ }
\end{eqnarray*}
The  simple
roots are  
\begin{equation*}
\alpha _{1}=e_{1}-e_{2},\text{ }\alpha _{2}=e_{2}-\zeta_{1}\text{ }
\end{equation*}%
In extending  the Lie super algebra $sl(2,1)$  to a twisted affine structure we assign to each generator 
a loop index $n$ which may be either, integer $n\in Z$ or semi-integer $n\in Z+1/2$.
Let  $Q= 2d + \h h_1$ be the grading operator and 
consider  
\begin{equation*}
E_{\pm }=h_1^{(1/2)} + 2h_2^{(1/2)} -\left( E_{\alpha _{1}}^{(0)}+E_{-\alpha _{1}}^{(1)}\right) 
\end{equation*}
When considering  the affine structure  of the $sl(2,1)$ super Kac-Moody algebra we find that there are 
 two affine twisted {\it subalgebras}  solving  the locality conditions (\ref{SUSY-local conditions}), namely,
\begin{itemize}
\item $sl(2,1)_{[1]}^{(2)}$
\end{itemize}
\begin{eqnarray}
{\cal K}_{Bose} &=& \{ K_1^{(2n+1)} = -(E_{\a_1}^{(n)} +E_{-\a_1}^{(n+1)}), \quad K_2^{(2n+1)} = \l_2\cdot H^{(n+1/2)}\}, \nonu \\
{\cal M}_{Bose} &=& \{ M_1^{(2n+1)} = -E_{\a_1}^{(n)} + E_{-\a_1}^{(n+1)}, \qquad M_2^{(2n)} = h_1^{(n)}\},
\label{sub1a} 
\er
and 
\br
{\cal K}_{Fermi} = \{ F_1^{(2n+3/2)} &=& (E_{\a_1+\a_2}^{(n+1/2)} -E_{\a_2}^{(n+1)} )+(E_{-\a_1-\a_2}^{(n+1)} - E_{-\a_2}^{(n+1/2)} ), \nonu \\
F_2^{(2n+1/2)} &=& -(E_{\a_1+\a_2}^{(n)} -E_{\a_2}^{(n+1/2)} )+(E_{-\a_1-\a_2}^{(n+1/2)} - E_{-\a_2}^{(n)} )\}, \nonu \\
{\cal M}_{Fermi} = \{ G_1^{(2n+1/2)} &=& ( E_{\a_1+\a_2}^{(n)} + E_{\a_2}^{(n+1/2)})+( E_{-\a_1-\a_2}^{(n+1/2)}+ E_{-\a_2}^{(n)}), \nonu \\
G_2^{(2n+3/2)} &=& -(E_{\a_1+\a_2}^{(n+1/2)}   + E_{\a_2}^{(n+1)} )+( E_{-\a_1-\a_2}^{(n+1 )} +E_{-\a_2}^{(n+1/2)}  )\}, \nonu \\
\label{sub1b}
\end{eqnarray}

\begin{itemize}
\item $sl(2,1)_{[2]}^{(2)}$
\end{itemize}
\begin{eqnarray}
{\cal K}_{Bose} &=& \{ K_1^{(2n+1)} = -(E_{\a_1}^{(n)} +E_{-\a_1}^{(n+1)}), \quad K_2^{(2n+1)} = \l_2\cdot H^{(n+1/2)}\}, \nonu \\
{\cal M}_{Bose} &=& \{ M_1^{(2n)} = -E_{\a_1}^{(n-1/2)} + E_{-\a_1}^{(n+1/2)}, \qquad M_2^{(2n+1)} = h_1^{(n+1/2)}\}, \nonu \\
\label{sub2a}
\er
and 
\br
{\cal K}_{Fermi} = \{ F_1^{(2n+1/2)} &=& (E_{\a_1+\a_2}^{(n)} - E_{\a_2}^{(n+1/2)} )+(E_{-\a_1-\a_2}^{(n+1/2)}- E_{-\a_2}^{(n)}  ), \nonu \\
F_2^{(2n+3/2)} &=& -( E_{\a_1+\a_2}^{(n+1/2)}-E_{\a_2}^{(n+1)})  + (E_{-\a_2}^{(n+1/2)}-  E_{-\a_1-\a_2}^{(n+1)} )\}, \nonu \\
{\cal M}_{Fermi} = \{ G_1^{(2n+1/2)} &=& ( E_{\a_1+\a_2}^{(n)} + E_{\a_2}^{(n+1/2)})+( E_{-\a_1-\a_2}^{(n+1/2)}+ E_{-\a_2}^{(n)}  ), \nonu \\
G_2^{(2n+3/2)} &=&  -( E_{\a_1+\a_2}^{(n+1/2)}  +E_{\a_2}^{(n+1)}) +(E_{-\a_1-\a_2}^{(n+1)}  +E_{-\a_2}^{(n+1/2)} )\}, \nonu \\
\label{sub2b}
\end{eqnarray}
The algebra  of such operators was given in the appendix A of ref \cite{n2}.


\subsection{ The sl(2,2) superalgebra.}
This algebra can be represented by 8 Bosonic, $\left\{ h_{1},\text{ }h_{2},\text{\ }h_{3},\text{ }E_{\pm \alpha
_{1}},\text{ }E_{\pm \alpha _{3}} \right\}$ and 8 Fermionic,  $\left\{ E_{\pm \alpha _{2}},\text{ }E_{\pm (\alpha _{1}+\alpha _{2})}\text{, }%
E_{\pm (\alpha _{2}+\alpha _{3})},\text{ }E_{\pm (\alpha _{1}+\alpha _{2}+\alpha _{3}})\right\}$
 generators.
The simple roots are
\begin{equation*}
\alpha _{1}=e_{1}-e_{2},\text{ }\alpha _{2}=e_{2}-\zeta_{1}\text{ , }\alpha
_{3}=\zeta_{1}-\zeta_{2}.
\end{equation*}
The principal grading for the $\widehat{ sl} (2|2)$ algebra
is defined in terms of the operator
\be
Q= d + \h (h_1+h_3)
\label{gradop}
\ee
and the grade one semisimple element $E$ is chosen as
\be
E_+ = (E_{\a_1}^{(0)} + E_{-\a_1}^{(2)}) + 
(E_{\a_3}^{(2)} + E_{-\a_3}^{(0)}) + I^{(1)} 
\label{esl22}
\ee
where is the identity, $I = h_1 +2h_2+h_3$.
The odd (fermionic) part of the kernel of $E_+$ of grade $n+1/2$ consists of
$ {\cal K}_{Fermi}= \{ f_i^{(n+\frac{1}{2})} , i=1,{\ldots} ,4, \;\;\;
n \in {\mathbb Z} \}$,
with
\br
f_1^{(n+\h)} &=&  (E_{\a_1+\a_2}^{(n-\h)} + E_{-\a_1-\a_2}^{(n+\frac{3}{2})}) + 
(E_{\a_2+\a_3}^{(n+\frac{3}{2})} + E_{-\a_2-\a_3}^{(n-\h)}), \nonu \\
f_2^{(n+\h)} &=&  (-E_{\a_1+\a_2}^{(n-\h)} + E_{-\a_1-\a_2}^{(n+\frac{3}{2})}) + 
(-E_{\a_2+\a_3}^{(n+\frac{3}{2})} + E_{-\a_2-\a_3}^{(n-\h)}), \nonu \\
f_3^{(n+\h)} &=&  (E_{\a_1+\a_2+\a_3}^{(n+\h)} + E_{-\a_1-\a_2-\a_3}^{(n+\h)}) + 
(E_{\a_2}^{(n+\h)} + E_{-\a_2}^{(n+\h)}), 
\nonu \\
f_4^{(n+\h)} &=&  (-E_{\a_1+\a_2+\a_3}^{(n+\h)} + E_{-\a_1-\a_2-\a_3}^{(n+\h)}) + 
(-E_{\a_2}^{(n+\h)} + E_{-\a_2}^{(n+\h)}), 
\nonu \\
\er
while the bosonic part 
$ {\cal K}_{Bose}= \{ K_i^{(n)} , i=1,{\ldots} ,3,\;\;\;
n \in {\mathbb Z} \} $ of the kernel ${\cal K}$ of grade $n$
contains
\[ K_1^{(n)} = E_{\a_1}^{(n-1)} + E_{-\a_1}^{(n+1)},\;\;\;
K_2^{(n)} = E_{\a_3}^{(n+1)} + E_{-\a_3}^{(n-1)},\;\;\;
K_3^{(n)}  = \l^n I \,.
\]
The fermionic part of the image ${\cal M}_{Fermi}$ of $E_+$ consists of
\be
{\cal M}_{Fermi}= \{ g_i^{(n+\frac{1}{2})} , i=1,{\ldots} ,4, \;\;\;
n \in {\mathbb Z} \}
\label{cmg22}
\ee
where
\br
g_1^{(n+\h)} &=&  (E_{\a_1+\a_2}^{(n-\h)} + E_{-\a_1-\a_2}^{(n+\frac{3}{2})}) - 
(E_{\a_2+\a_3}^{(n+\frac{3}{2})} + E_{-\a_2-\a_3}^{(n-\h)}), \nonu \\
g_2^{(n+\h)} &=&  (-E_{\a_1+\a_2}^{(n-\h)} + E_{-\a_1-\a_2}^{(n+\frac{3}{2})}) +
(E_{\a_2+\a_3}^{(n+\frac{3}{2})} - E_{-\a_2-\a_3}^{(n-\h)}), \nonu \\
g_3^{(n+\h)} &=&  (E_{\a_1+\a_2+\a_3}^{(n+\h)} + E_{-\a_1-\a_2-\a_3}^{(n+\h)}) -
 (E_{\a_2}^{(n+\h)} + E_{-\a_2}^{(n+\h)}), 
\nonu \\
g_4^{(n+\h)} &=&  (-E_{\a_1+\a_2+\a_3}^{(n+\h)} + E_{-\a_1-\a_2-\a_3}^{(n+\h)}) + 
(E_{\a_2}^{(n+\h)} - E_{-\a_2}^{(n+\h)}), 
\nonu \\
\label{gg}
\er
There are four bosonic generators  
\begin{alignat}{2}
M_1^{(n)} &= h_1^{(n)}, &\qquad  M_2^{(n)} &= -E_{\a_1}^{(n-1)} + E_{-\a_1}^{(n+1)}, \nonu \\
M_3^{(n)} &=h_3^{(n)}, &\qquad 
M_4^{(n)} &= -E_{\a_3}^{(n+1)} + E_{-\a_3}^{(n-1)}, \nonu 
\end{alignat}
in the image
${\cal M}$ of $E_+$. Note that $M_1^{(n)}$ and $M_3^{(n)}$ are in 
the Cartan subalgebra. 
This algebra can be represented by $4\times 4$ block matrices 
and their structure  is given in \cite{n2}. 
Define
\br
F_1^{(n+1/2)} &=& {{1}\over {\sqrt{2}}}(f_1^{(n+1/2)} + f_3^{(n+1/2)}), \quad F_2^{(n+1/2)} = {{1}\over {\sqrt{2}}}(f_2^{(n+1/2)} + f_4^{(n+1/2)})\nonu \\
F_3^{(n+1/2)} &=& {{1}\over {\sqrt{2}}}(f_1^{(n+1/2)} - f_3^{(n+1/2)}), \quad F_4^{(n+1/2)} = {{1}\over {\sqrt{2}}}(f_2^{(n+1/2)} - f_4^{(n+1/2)})\nonu \\
\label{f}
\er 
\br
G_1^{(n+1/2)} &=& {{1}\over {\sqrt{2}}}(g_1^{(n+1/2)} + g_3^{(n+1/2)}), \quad G_2^{(n+1/2)} = {{1}\over {\sqrt{2}}}(g_2^{(n+1/2)} + g_4^{(n+1/2)})\nonu \\
G_3^{(n+1/2)} &=& {{1}\over {\sqrt{2}}}(g_1^{(n+1/2)} - g_3^{(n+1/2)}), \quad G_4^{(n+1/2)} = {{1}\over {\sqrt{2}}}(g_2^{(n+1/2)} - g_4^{(n+1/2)})\nonu \\
\label{g}
\er
The four subalgebras which solves the locallity conditions are defined by:
\begin{eqnarray}
sl(2,2)_{[1]}^{(2)} &=&\left\{ 
\begin{array}{c}
K_{1}^{2n+1},K_{2}^{2n+1},\text{ }K_{3}^{2n+1},\text{ }M_{1}^{2n},\text{ }
\\ 
M_{2}^{2n+1},M_{3}^{2n},M_{4}^{2n+1},\text{ }F_{1}^{2n+3/2},\text{ } \\ 
F_{2}^{2n+1/2},F_{3}^{2n+3/2},F_{4}^{2n+1/2},\text{ }G_{1}^{2n+1/2},\text{ }
\\ 
G_{2}^{2n+3/2},G_{3}^{2n+1/2},G_{4}^{2n+3/2}%
\end{array}%
\right\}  \label{Ia} \\
sl(2,2)_{[2]}^{(2)} &=&\left\{ 
\begin{array}{c}
K_{1}^{2n+1},K_{2}^{2n+1},\text{ }K_{3}^{2n+1},\text{ }M_{1}^{2n+1},\text{ }
\\ 
M_{2}^{2n},M_{3}^{2n+1},M_{4}^{2n},\text{ }F_{1}^{2n+1/2},\text{ } \\ 
F_{2}^{2n+3/2},F_{3}^{2n+1/2},F_{4}^{2n+3/2},\text{ }G_{1}^{2n+1/2},\text{ }
\\ 
G_{2}^{2n+3/2},G_{3}^{2n+1/2},G_{4}^{2n+3/2}%
\end{array}%
\right\}  \label{IIa} \\
sl(2,2)_{[3]}^{(2)} &=&\left\{ 
\begin{array}{c}
K_{1}^{2n+1},K_{2}^{2n+1},\text{ }K_{3}^{2n+1},\text{ }M_{1}^{2n},\text{ }
\\ 
M_{2}^{2n+1},M_{3}^{2n+1},M_{4}^{2n},\text{ }F_{1}^{2n+1/2},\text{ } \\ 
F_{2}^{2n+3/2},F_{3}^{2n+3/2},F_{4}^{2n+1/2},\text{ }G_{1}^{2n+1/2},\text{ }
\\ 
G_{2}^{2n+3/2},G_{3}^{2n+3/2},G_{4}^{2n+1/2}%
\end{array}%
\right\}  \label{IIIa} \\
sl(2,2)_{[4]}^{(2)} &=&\left\{ 
\begin{array}{c}
K_{1}^{2n+1},K_{2}^{2n+1},\text{ }K_{3}^{2n+1},\text{ }M_{1}^{2n+1},\text{ }
\\ 
M_{2}^{2n},M_{3}^{2n},M_{4}^{2n+1},\text{ }F_{1}^{2n+3/2}, \\ 
\text{ }F_{2}^{2n+1/2},F_{3}^{2n+1/2},F_{4}^{2n+3/2},\text{ }G_{1}^{2n+1/2},%
\text{ } \\ 
G_{2}^{2n+3/2},G_{3}^{2n+3/2},G_{4}^{2n+1/2}%
\end{array}%
\right\}  \label{IVa}
\end{eqnarray}

\section{Appendix C:  Vector, Scalar and Pseudo Scalar Currents } 
Here we give explicit relations of 
vector, scalar and pseudo scalar currents   in terms of Fermi fields components parametrizing $W_{\pm 1/2}$.
\br
\bar {\Psi}_D\Psi_D &=& \psi_1 \bar \psi_3 + \bar \psi_1 \psi_3\nonu \\
\bar {\Phi}_D\Phi_D &=& \psi_2 \bar \psi_4 + \bar \psi_2 \psi_4
\er
\br
\bar {\Psi}_D \g^5 \Psi_D = -\psi_1 \bar \psi_3 + \bar \psi_1 \psi_3\nonu \\
\bar {\Phi}_D \g^5 \Phi_D = -\psi_2 \bar \psi_4 + \bar \psi_2 \psi_4
\er

\br
\bar {\Psi}_D \g^0 \Psi_D = -i\psi_1  \psi_3 +i \bar \psi_1 \bar \psi_3\nonu \\
\bar {\Phi}_D \g^0 \Phi_D = -i\psi_2  \psi_4 + i\bar \psi_2 \bar \psi_4 \nonu \\
\bar {\Psi}_D \g^1 \Psi_D = i\psi_1  \psi_3 -i \bar \psi_1 \bar \psi_3\nonu \\
\bar {\Phi}_D \g^1 \Phi_D = -i\psi_2  \psi_4 + i\bar \psi_2 \bar \psi_4
\er


\end{document}